\documentclass[prd,article,nofootinbib,twocolumn,preprintnumbers]{revtex4}

\pdfoutput=1
\usepackage[utf8x]{inputenc}
\usepackage{amsthm}
\usepackage{amsmath}
\usepackage{amsfonts}
\usepackage{todonotes}
\usepackage{amssymb}
\usepackage{graphicx,amsfonts,color,comment,amsmath,hyperref,float}
\usepackage{bm}
\usepackage{cancel}
\usepackage{epstopdf}
\usepackage{bbold}

\usepackage{xspace}
\usepackage{axodraw4j}
\usepackage{pstricks}
\usepackage{color}

\newcommand{\beq}{\begin{eqnarray}}
\newcommand{\eeq}{\end{eqnarray}}
\newcommand{\GeV}{\,\text{GeV}}
\newcommand{\TeV}{\,\text{TeV}}

\newcommand{\ifb}{{\,\rm fb}^{-1}}

\newcommand{\iab}{{\,\rm ab}^{-1}}

\newcommand{\ord}{\mathcal{O}}
\newcommand{\mca}{\mathcal{A}}
\newcommand{\imag}{\mathrm{Im}}
\newcommand{\gt}{\widetilde{g}}

\DeclareMathOperator{\tr}{Tr}

\def\mg5{\textsc{MadGraph5\_aMC\@NLO}}
\def\Madgraph{\textsc{MadGraph5\_aMC\@NLO}}
\def\Pythia8{\textsc{Pythia8}}

\def\eq#1{{eq.~(\ref{#1})}}
\def\eqs#1#2{{eqs.~(\ref{#1})--(\ref{#2})}}
\def\fig#1{{fig.~(\ref{#1})}}
\def\sec#1{{sec.~(\ref{#1})}}
\def\fig#1{{fig.~(\ref{#1})}}
\def\tab#1{{tab.~(\ref{#1})}}

\def\app#1{{app.~(\ref{#1})}}

\newcommand{\FeynRules}{F\protect\scalebox{0.8}{EYN}R\protect\scalebox{0.8}{ULES}\xspace}

\newcommand{\FCC}{FCC\xspace}
\newcommand{\MEPS}{M\scalebox{0.8}{E}P\scalebox{0.8}{S}@LO\xspace}
\newcommand{\Rivet}{R\protect\scalebox{0.8}{IVET}\xspace}
\newcommand{\Sherpa}{S\protect\scalebox{0.8}{HERPA}\xspace}
\newcommand{\CSS}{C\protect\scalebox{0.8}{SS}\xspace}
\newcommand{\Comix}{C\protect\scalebox{0.8}{OMIX}\xspace}
\newcommand{\LHC}{LHC\xspace}

\begin{document}

\title{Implications of Vector Boson Scattering Unitarity in Composite Higgs Models}


\author{Diogo Buarque Franzosi
and Piero Ferrarese
\\
II. Physikalisches Institut, Universit\"at G\"ottingen, \\Friedrich-Hund-Platz 1, 37077 G\"ottingen, Germany 
\\
{\tt dbuarqu@gwdg.de, piero.ferrarese@phys.uni-goettingen.de} \smallskip}

\begin{abstract}

The strong nature of Composite Higgs models manifests at high energies through the growing behavior of the scattering amplitudes of longitudinally polarized weak bosons 
that leads to the formation of composite resonances as well as non-resonant strong effects.
In this work the unitarity of these scattering amplitudes
is used as a tool to assess the profile of the composite spectrum of the theory, including non-resonant enhancements, vector resonances and the CP-even scalar excitation.
These three signatures are then studied in realistic scattering processes at hadron colliders, aiming to estimate the potential to exclude dynamically motivated scenarios of Composite Higgs models.

\end{abstract}
 
\maketitle

\section{Introduction}
\label{sec:introduction}

The mechanism that spontaneously breaks Electroweak (EW) Symmetry is still unveiled.
With the discovery of the Brout-Englert-Higgs (BEH) boson~\cite{Englert:1964et,Higgs:1964ia} in 2012 by the ATLAS and CMS collaborations~\cite{Aad:2012tfa,Chatrchyan:2012xdj,Chatrchyan:2013lba}, a new key piece of the puzzle came into play, with nonetheless no completion of a greater picture on the horizon. 
What is missing is a \emph{natural} explanation for the generation of the Higgs potential, without enormous fine-tuning and preferably from a dynamical origin without \emph{ad-hoc} terms as in the Standard Model (SM).
Both the longitudinal weak bosons and the BEH boson are part of the spontaneous EW Symmetry Breaking (EWSB) sector thus their thorough investigation at the Large Hadron Collider (LHC) and future colliders will eventually shed more light on the topic. 

Composite Higgs (CH) models  are among the most promising candidates to address some of SM weaknesses, dynamically generating the EW scale  through a vacuum condensate and at the same time explaining the mass gap between the Higgs boson and the other composite states by the identification of the Higgs with one of the pseudo-Nambu Goldstone boson (NGB) of the underlying global symmetry breaking ~\cite{Kaplan:1983fs,Kaplan:1983sm,Georgi:1984af}\footnote{For reviews of the more recent developments in the CH literature we refer the reader to Refs.~\cite{Bellazzini:2014yua,Panico:2015jxa}.}.
In CH models the fermionic condensate generating the EW scale is misaligned with respect to the vacuum that breaks the EW group, thus the acquirement of a vacuum expectation value (\emph{vev}) by the fermionic condensate creates a hierarchy between the NGB decay constant $f$ and the EW scale $v$, in the form,
\begin{equation}
v=f \sin\theta\,,
\label{eq:f2v}
\end{equation}
where $\theta$ is the misalignment angle.

A striking evidence of new strong dynamics at high scales  is the presence of strong Vector Boson Scattering (VBS)~\cite{Veltman:1976rt,Veltman:1977fy,Lee:1977yc,Lee:1977eg,Passarino:1985ax,Passarino:1990hk}, or more generally strong Goldstone Boson Scattering (GBS), including physical pseudo-NGBs (the Higgs boson itself and others in non-minimal CH realisations) and longitudinal VBS, which are related to the GBS by the Goldstone Boson Equivalence Theorem (GBET)~\cite{Cornwall:1974km}.
The strong nature of the NGBs in CH models manifests itself in GBS through the  miscancellation of Feynman diagrams and the divergent behavior of the scattering amplitudes according to the Low Energy Theorems (LET)\cite{Weinberg:1966kf}
\begin{equation}
\mathcal{A}(\pi\pi\to \pi\pi)\sim \frac{s}{f^2} = \frac{s}{v^2}\sin^2\theta \,,
\label{eq:letsimp}
\end{equation}
with $\pi$ a NGB and $s$ the Mandelstam variable.
This is in contrast with the well behaved amplitudes of the SM.
The growing behavior of GBS amplitudes must be eventually controlled by strong effects at high energies, either in the form of broad \emph{continuum} enhancements or in the form of composite resonances, saturating unitarity similarly to what happens in hadron physics.
In this case it remains to be quantified the possibility to probe such high scales at the LHC or if a higher energy machine, such as the FCC 100 TeV collider will be necessary.

To tackle this problem the first question that arises is how to estimate the actual scale of resonance formation or strong \emph{continuum} effects. 
The unitarity of $2\to 2$ GBS amplitudes computed at fixed order in perturbation theory has been used as a tool to set limits on the scale of new physics  or strong interactions, and on the mass of a heavy Higgs boson~\cite{Lee:1977eg,PASSARINO199031,Chanowitz:1985hj}.
We will pursue this idea in the context of CH models. 

Through the analysis of the GBS amplitudes and under the guidance of unitarity principles we will set limits on the scale of resonance formation, in particular in the scalar channel which is only poorly described by lattice calculations. We will show that near the scale of leading-order (LO) unitarity violation the \emph{continuum} of LET dominates the scattering amplitudes and prevents the formation of Breit-Wigner resonances.
We will also argue that we can not only set constraints on the masses of resonances, but also on their couplings, if we assume basic criteria of saturation of unitarity and analyticity provided  by the Inverse Amplitude Method  (IAM) of unitarization.

Following our assessment of resonance profiles via the study of unitarization of GBS amplitudes, we will estimate the potential to observe strong effects in realistic observables, whether resonances or strong \emph{continuum} effects dominate the amplitudes. 
We will analyze the production cross sections of heavy vector resonances through weak boson fusion (VBF) and Drell-Yan (DY) and non-resonant and scalar-resonant scenarios of strong VBS in $pp\to jjZZ\to jj4\ell$ channel.

Although our results are general and can be extended to other scenarios, we will use as template  the Fundamental Minimal CH Model (FMCHM)~\cite{Galloway:2010bp,Cacciapaglia:2014uja}, whose description and effective construction will be given in \sec{sec:effmodel}.
After the description of our CH template we present in \sec{sec:unitarity}  a detailed analysis of the GBS amplitudes, their unitarity constraints and implications for the spectrum of composite states.
In the light of the results from \sec{sec:unitarity},  we will study in \sec{sec:experimental} the possibility of observing signals of strong VBS and heavy vector production at the LHC and a future 100 TeV collider.

\section{FMCHM}
\label{sec:effmodel}

The FMCHM is based on the coset SU(4)/Sp(4), which has been studied in several previous works~\cite{Duan:2000dy,Katz:2005au,Lodone:2008yy,Gripaios:2009pe}, 
and is the simplest global symmetry breaking pattern which can be realized  in terms of an underlying fermionic gauge theory\footnote{The minimal CH model, SO(5)/SO(4), can be realized with the inclusion of 4-fermion operators~\cite{vonGersdorff:2015fta}.}. 
The simplest underlying theory realizing this symmetry breaking is based on the SU(2) gauge theory with two Dirac fermions
transforming according to the fundamental representation of the  gauge group~\cite{Ryttov:2008xe,Galloway:2010bp,Cacciapaglia:2014uja}.
This UV completion has been studied as CH model in Ref.~\cite{Galloway:2010bp}, and unified in a \emph{pure} Technicolor scenario, in which the Higgs boson is identified  with a scalar excitation \emph{techni}-$\sigma$ in Ref.~\cite{Cacciapaglia:2014uja}.  

In SU(4)/Sp(4) a general parametrization of the vacuum is given by
\begin{equation}
\Sigma_0 = \cos\theta\Sigma_B+\sin\theta\Sigma_H\,,
\end{equation}
where we can choose to embed the EW group SU(2)$\times$U(1) in such a way that $\Sigma_H$ fully breaks EW symmetry while $\Sigma_B$ does not.
The interactions between the strong sector and the top-quark tend to favor the vacuum that fully breaks EW symmetry~\cite{Agashe:2004rs}, but other forces, such as gauge interactions and explicit \emph{techni}-quark masses, play a role in stabilizing the potential in an intermediate value of the misalignment angle $\theta$.

The mechanism to generate the top-quark mass originates also from these interactions with the strong sector.
In the  extended technicolor (ETC) description the top mass is generated via 4-fermion operators bilinear in the top quark, discussed for the $SU(4)/Sp(4)$ coset in Refs.~\cite{Galloway:2010bp,Cacciapaglia:2014uja}. These bilinear 4-fermion interactions may arise from the exchange of heavy spin-1 bosons~\cite{Eichten:1979ah,Dimopoulos:1979es} or heavy scalars~\cite{'tHooft:1979bh} external to the strongly interacting sector considered here. A recent explicit example employing a chiral gauge theory is provided in \cite{Cacciapaglia:2015yra}.

Another possibility is provided by fermion partial compositeness~\cite{Kaplan:1991dc}, a mechanism that requires the presence of fermionic bound states that mix linearly to the elementary fermion fields. 
Partial compositeness has been explored for SU(4)/Sp(4) coset in~\cite{Ferretti:2013kya,Barnard:2013zea,Ferretti:2016upr}. In this case, at least another pair of \emph{techni-}fermions charged under the QCD gauge group must be incorporated to allow the formation of a composite state with the same quantum numbers of the top quark. 
The top partner may be important to stabilize the Higgs potential, in that case it is expected to be parametrically lighter than the typical resonance scale~\cite{Matsedonskyi:2012ym}, but other spurions like mass terms for the underlying fermions can be used as stabilizer, and the top-partner may be heavy and irrelevant for the  phenomenology presented here. Also other NGBs are expected to pop up in this configuration; however, a Dirac mass term to the new fermions is allowed and avoid these additional states in the spectrum~\cite{Barnard:2013zea}.


In both mechanisms mentioned above, the model may be constrained by flavour observables, especially in the form of Flavour Changing Neutral Currents (FCNC) induced by four-fermion operators at the flavour scale.
One way under consideration to avoid such constraints relies  on the presence of a Conformal Theory in the UV, that generates large anomalous dimensions to enhance the condensate as opposed to SM 4-fermion interactions and FCNC. Unfortunately, recent results indicate that obtaining large anomalous dimensions is challenging, both for scalar operators ~\cite{Rattazzi:2008pe,Rychkov:2009ij,Rattazzi:2010yc}, needed for ETC-like masses, as well as for fermionic ones, needed in partial compositeness~\cite{Pica:2016rmv,Vecchi:2016whd}. 
A full solution to the flavor hierarchy is thus still missing.

Independently of the specific mechanism to generate top mass, the terms of the effective potential originate from the same strong sector and have the same mass scale. Typically this fact implies that $\sin\theta$ is a good parametrization of \emph{fine-tuning} in the model~\cite{Ferretti:2016upr,Cacciapaglia:2014uja} and $\theta$ is thus naturally not so small. 
On the other hand, large angles are not favored by data due to deviations of the Higgs couplings from the SM predictions, which upset EW precision observables (EWPO) resulting in an upper bound~\cite{Arbey:2015exa}
\begin{equation}
\sin\theta\lesssim 0.2 \quad \mbox{(EWPO)}\,\footnote{This limit depends mildly on the fermion content and dynamics of the underlying theory but is dominated by Higgs coupling modification. It can also be alleviated by cancellations from other composite states~\cite{Franzosi:2016aoo}.}.
\label{eq:ewpo-th}
\end{equation}

We use the Callan-Coleman-Wess-Zumino (CCWZ) construction~\cite{Coleman:1969sm,Callan:1969sn} to write the effective Lagrangian.
The lowest dimension ($d=2$) term is given by 
\begin{equation}
\mathcal{L}_2 = \frac{1}{2} f^2 \langle x_{\mu} x^{\mu} \rangle ,
\label{eq:LagLO}
\end{equation}
where $x_\mu$ is the projection of the Maurer-Cartan form and contains the 5 NGBs (see \app{app:ccwz} for conventions and more details). $\langle  A \rangle$  is the trace of the matrix $A$.

In order to analyze unitarity it is imperative to include higher order terms due to the strong relation between perturbativity and unitarity. Since the CCWZ Lagrangian is an effective \emph{non-renormalizable} theory, each order in the perturbation expansion has to be accompanied by a tower of higher dimension operators in order to carry out the renormalization program.
The $d=4$ Lagrangian is given by
\begin{eqnarray}
\mathcal{L}_4 &=& L_0\langle x^{\mu}x^{\nu}x_{\mu}x_{\nu}\rangle + L_1 \langle x^{\mu}x_{\mu}\rangle \langle x^{\nu}x_{\nu} \rangle \nonumber\\
              &+& L_2 \langle x^{\mu}x^{\nu} \rangle \langle x_{\mu}x_{\nu} \rangle + L_3 \langle x^{\mu}x_{\mu}x^{\nu}x_{\nu}\rangle \,.
\label{eq:ldim6}
\end{eqnarray}

\subsection{Vector Resonances}
\label{sec:vector}

The composite vector resonances in FMCHM have been studied in Ref.~\cite{Franzosi:2016aoo} making use of the hidden local symmetry (HLS) approach \cite{Bando:1987br}. 
In the FMCHM a vast spectrum of 15 heavy composite vector resonances is expected, with very peculiar phenomenology. They can be associated with the broken generators $Y_a$ and the unbroken ones $V_a$,
\begin{equation}
\bm{\mathcal{F}}_\mu = \bm{\mathcal{V}}_\mu+\bm{\mathcal{A}}_\mu = \sum_{a=1}^{10} {\cal V}_\mu^a V_a + \sum_{a=1}^{5} {\cal A}_\mu^a Y_a,
\label{eq:vectors}
\end{equation}
 forming a {\bf 10} and a {\bf 5} multiplet of Sp(4). 
The lowest dimension Lagrangian is given by (see \app{app:ccwz} for details and conventions)
\begin{eqnarray}
{\cal L}_v 
&=&-\frac{1}{2\gt^2}\ \langle \bm{\mathcal{F}}_{\mu\nu} \bm{\mathcal{F}}^{\mu\nu}\rangle +\frac{1}{2}f_0^2\ \langle x_{0\mu} x_0^\mu \rangle \nonumber\\
       &+& \frac{1}{2}f_1^2\ \langle x_{1\mu} x_1^\mu \rangle +r f_{1}^2\ \langle x_{0\mu} K x_1^\mu K^\dagger \rangle \nonumber \\
       &+& \frac{1}{2} f_K^2\ \langle { D}^\mu K\ { D}_\mu K^\dagger \rangle \ .
\label{eq:veclag}
\end{eqnarray} 
$ \bm{\mathcal{F}}_{\mu\nu}$ is the field strength tensor of $\bm{\mathcal{F}}_\mu$. 
The EW \emph{vev} is
\beq \label{eq:v2}
v^2 = \left( f_0^2 - r^2 f_1^2 \right) \sin^2 \theta = f_\pi^2 \sin^2 \theta \,,
\eeq
where $f_\pi = \sqrt{f_0^2 - r^2 f_1^2}$. 
We neglect possible direct couplings of $\bm{\mathcal{F}}_\mu$ to fermions, which are generated in our set-up only through the mixing with EW gauge bosons. 

The masses of $\bm{\mathcal{V}}_\mu$ and $\bm{\mathcal{A}}_\mu$ (without EW interactions) are given respectively by
\beq
M_V\equiv\frac{\widetilde{g}f_K}{\sqrt{2}} \quad \text{   and   }  \quad M_{\bm{\mathcal{A}}}\equiv\frac{\widetilde{g}f_1}{\sqrt{2}} \,.
\eeq
These masses have been estimated with lattice calculations for the FMCHM SU(2) gauge theory with 2 Dirac fermions, $M_V=3.2(5)\TeV/\sin\theta$ and $M_{\bm{\mathcal{A}}}=3.6(9)\TeV/\sin\theta$~\cite{Arthur2016a}.

Once the masses are fixed there are 2 extra free parameters which were not computed from first principles: $\gt$ and $r$.  
These parameters basically determine the branching ratios into fermions or bosons. If $r= 1$ the fermion decays dominate, once $|r-1|\gtrsim 0.1$ the diboson decays dominate.

We evaluate now the trilinear couplings between heavy vectors and the GBs, which will be important for our analysis of GBS. They come from the $f_K$ term in \eq{eq:veclag}. Only couplings to $\bm{\mathcal{V}}_\mu$ are generated, which can be expressed as
\begin{eqnarray}
\pi_a(p_1) \pi_b(p_2) {\cal V}_\mu^c: & & \frac{2\gt f_K^2(1-r^2)}{f^2}\tr(Y^a Y^b V^c) (p_1 - p_2) \nonumber\\
                            &&= ig_V(p_1 - p_2) \Xi^{abc},
\label{eq:vtril}                            
\end{eqnarray}
where 
\begin{equation}
g_V=-\frac{M_V}{2f}a_V = -\frac{M_V^2(1-r^2)}{\sqrt{2}\tilde{g}f^2}\,,
\label{eq:vectcoup}
\end{equation} 
and $\Xi^{abc}=1$ for $(c,a,b)$= (1,3,2), (2,3,1), (3,1,2), (4,1,4), (5,2,4), (6,3,4), (7,5,3), (8,5,4), (9,5,1), (10,2,5), $\Xi^{abc}=-1$ by interchanging $a \leftrightarrow b$ above and $\Xi^{abc}=0$ otherwise.
The couplings in terms of charge eigenstates are shown in \eqs{eq:vgcoup1}{eq:vgcoup2}.

\subsection{Scalar isosinglet $\sigma$}
\label{sec:sigma}

Additional scalars are a common feature in composite extensions of the SM, see \emph{e.g.}~\cite{Cacciapaglia:2015eqa,Ferretti:2016upr}.
The scalar singlet $\sigma$ can be incorporated in a simple general way
\begin{equation}
\mathcal{L}_\sigma = \frac{1}{2}\kappa (\sigma/f) f ^{2} \langle x_{\mu}x^{\mu} \rangle 
 + \frac{1}{2} \partial_\mu \sigma \partial^\mu \sigma -\frac{1}{2} M_\sigma^2\sigma^2 \,,
 \label{eq:sigma}
\end{equation}
with $\kappa(\sigma/f)=1+\kappa'\sigma/f+\kappa''\sigma^2/(2f^2)+\cdots$.
The potential (which must be added to $\mathcal{L}_\sigma$) generates a tadpole term that drives the \emph{vev} to $\sigma$. In addition, it also generates a mixing term with the Higgs boson, $h$. These effects are however small when a very heavy scalar is considered.  Mixing between $h$ and $\sigma$ is, for small $\theta$,  approximately $\alpha\sim \frac{2m_h^2}{m_\sigma^2}$\cite{Arbey:2015exa}.
For a $M_\sigma\gtrsim 5\TeV$, $\alpha\lesssim 0.00125$ is very small and will be neglected in the following analysis.


The relevant parameters are here $M_\sigma$ and $\kappa'$. The lattice prediction for the SU(2) gauge theory with 2 Dirac fermions has large uncertainty $M_\sigma=4.7(2.6)\TeV/\sin\theta$\cite{Arthur2016a}. We will see that unitarity of VBS provides more stringent limits on the parameters of this state.
The trilinear couplings between $\sigma$ and the NGBs read 
\begin{equation}
\sigma \pi_a(p_1)\pi_b(p_2) : -2i\frac{g_\sigma}{f}p_1\cdot p_2\,\delta_{ab}
\end{equation}
with $g_\sigma=\kappa'/2$.

\section{Unitarity Implications}
\label{sec:unitarity}

In this section we will analyze the $2\to 2$ scattering of NGBs and by requiring the fulfillment of unitarity condition we will make predictions for the model parameters and spectra discussed above.
We use the FMCHM as template but our results can be easily generalized. 

Let us consider a $\pi\pi\to \pi\pi$ elastic scalar scattering  amplitude, $\mathcal{A}(s,t)$, with $s$ and $t$ the Mandelstam variables, and expand it in partial waves:
\begin{eqnarray}
\mathcal{A} (s,t)&=&32\pi\sum_{J=0}^\infty a_J(s)(2J+1)P_J(\cos\theta),\nonumber\\
a_{J}(s) &=& \frac{1}{32 \pi\, s} \int_{-s}^{0} \mathrm{d} t \mathcal{A} (s,t) P_J (x)\,,
\label{eq:pw}
\end{eqnarray}
where $x$ is the cosine of the  scattering angle and $P_J(x)$ are the Legendre polynomials. 
In this basis elastic partial wave unitarity condition reads
\begin{equation}
\mathrm{Im} a_J(s)= |a_J(s)|^2\,.
\label{eq:pwunit}
\end{equation}

In order to force elasticity in the NGB sector
it is customary to expand the amplitudes in definite conserved quantum numbers before expanding them in partial waves. 
In the chiral Lagrangian of pions the $\pi\pi$ scattering amplitudes can be expanded in the usual definite isospin $I$.
For SU(4)/Sp(4), just like isospin, we expect that specially at high energies Sp(4) is approximately unbroken and we can therefore expand the $2\to 2$ NGB scattering in definite multiplets of Sp(4), as
\begin{equation}
\bf 5\otimes 5 = 1\oplus 10\oplus 14\,,
\label{eq:decomposition}
\end{equation}
and assume they correspond to pure elastic channels, with no mixing among them. 
We note that the Higgs boson is also part of the NGB scattering.
The VBS topology can be seen as a special case of this scattering, with the longitudinal modes related to the eaten NGB $\pi^i$ (i=1,2,3) through the equivalence theorem.

Inelastic channels $\pi\pi\to X$ with $X$ being either $n>2$ NGBs, other composite states out of the NGB multiplet or SM particles have the effect of squeezing the Argand circle to lower radius,  
$\mathrm{Im} a_J(s)> |a_J(s)|^2$. To be more precise, we can define inelastic scattering amplitudes that fulfill (see \app{app:pw} for more details)
\begin{equation}
\sum_X|a_J^{X}(s)|^2\sqrt{1-M_X^2/s}\equiv \imag a_J(s) - |a_J(s)|^2\,,
\label{eq:pwinel}
\end{equation}
where 
$M_X^2$ is the sum of squared masses of all particles in the $X$ system and $\sum_X|a_J^{X}(s)|^2\sqrt{1-M_X^2/s}\leq 1/4$. Eq.\ref{eq:pwinel} goes back to \eq{eq:pwunit} in the purely elastic limit. Note that knowing the cross section of the inelastic channels one can get the exact unitarity condition and thus the scale of unitarity violation and strong effects accordingly.

Nevertheless, elasticity is a good approximation for energies below the threshold production of composite states other than the NGB. 
Inelastic channels with more than 2 NGB in the final state appear at higher order in chiral perturbation theory at $\ord \left( (s/(4\pi f)^2)^3 \right)$ while the production of SM fermions and transversal weak bosons is negligible at high energies~\cite{Lee:1977eg}.
Other composite states can open relevant inelastic channels beyond their mass thresholds, \emph{e.g.} $\sqrt{s}>2M_\sigma$ for the scalar; however, since we are interested in describing the lightest resonance in each \emph{isospin}-spin channel, these effects can be usually neglected, and we will comment in the text when otherwise.
%

Two interesting exceptions in which inelasticity could be relevant are the case of a relatively light top partner $T$ and the case of extra NGBs present in larger groups (like the ones required in partial compositeness scenario). 
In both cases, it is plausible that these states sit at scales higher than we consider here  (as discussed in the introduction).
%
It is nevertheless worth to comment on the possibility of a lighter top partner $T$, which opens inelastic channels via the process $\pi\pi\to T\bar{T}$ with $\pi\pi T$ couplings typically of order ${\cal O}(M_T/f)$~\cite{DeSimone:2012fs} and would affect our analysis. 
A thorough  analysis of the contribution of this process to partial waves is still missing, although in Refs.~\cite{Chanowitz:1978uj,Chanowitz:1978mv} the scale of strong interaction in $FF\to FF$ channel were estimated for a heavy fermion $F$.
The analysis of these processes will be presented elsewhere.

From the $d=2$ Lagrangian (\eq{eq:LagLO}) we get the LO amplitudes. 
We will consider only the leading spin, \emph{i.e.} the scalar $J=0$ for the singlet channel ${\bf 1}\equiv A$, $a_{A0}(s)$, the vector $J=1$ for the ${\bf 10}\equiv B$ representation  (since the $J=0$ amplitude vanishes), $a_{B1}(s)$, and the scalar for ${\bf 14}\equiv C$, $a_{C0}(s)$. 
The corresponding partial wave amplitudes at LO are:
\begin{eqnarray}
a^{(0)}_{A0}(s)&=&\frac{s}{16\pi f^2}\,, \label{eq:aI0}\\
a^{(0)}_{B1}(s)&=&\frac{s}{192\pi f^2}\,, \\
a^{(0)}_{C0}(s)&=&\frac{-s}{64\pi f^2}\,.
\end{eqnarray}

A real amplitude can never satisfy the unitarity condition, \eq{eq:pwunit}. The absorptive and imaginary part of the amplitude comes at first order at loop level, and fulfills the perturbative unitarity relation~\footnote{If no inelastic channels are open, otherwise further loop contribution will increase the imaginary part with $\imag\, a^{(1)}(s)>|a^{(0)}(s)|^2$.}
\begin{equation}
\imag\, a^{(1)}(s)=|a^{(0)}(s)|^2\,,
\end{equation}
where $a^{(1)}(s)$ is the correction from the effective energy expansion. 
As long as perturbativity is under control this relation is sufficient to avoid unitarity violation, what makes evident the relation between unitarity and perturbativity.
This observation allows us to define another criteria of unitarity, which is
\begin{equation}
|a(s)|<1\,.
\label{eq:unitviolation}
\end{equation}
Therefore, according to \eq{eq:aI0} unitarity is faded to be violated at energies
\begin{equation}
\sqrt{s}\gtrsim 4\sqrt{\pi}f\,.
\label{eq:lambdalo}
\end{equation}
Since from EWPO we expect $\sin\theta\lesssim 0.2$, we need to reach partonic energies of the order of $\sqrt{s}\lesssim 8 \TeV$ to observe strong VBS effects. Such energies could in principle be at the extreme corner of LHC potential, but it seems more feasible to be reached at a higher energy machine, such as a 100 TeV collider. Even for lower angles, \emph{e.g.} $\theta=0.1$, unitarity violation would take place around $\sqrt{s}\sim 16 \TeV$, which is within the reach of a 100 TeV machine.
If inelastic channels are open, the scale of strong effects are expected to be lower.

The next-to-leading order (NLO) correction to the partial wave amplitudes, which includes the tree level diagrams involving dimension-6 operators, \eq{eq:ldim6}, and one-loop diagrams, is given by~\cite{Bijnens:2011fm} 
\begin{widetext}
\begin{eqnarray}
a^{(1)}_{A0}(s)&=&\frac{s^2}{32 \pi f^4}  
 \left[\frac{1}{16\pi^2}\left(\frac{29}{12}+\frac{46}{18}\log \left(\frac{s}{\mu ^2}\right)+2\pi i\right)+
 \frac{2}{3}\widehat{L_A}(\mu)\right]\,, \\
a^{(1)}_{B1}(s)&=&\frac{s^2}{32 \pi f^4}  
 \left[\frac{1}{16\pi^2}\left(-\frac{35}{432}+\frac{1}{12}\log \left(\frac{s}{\mu ^2}\right)+\frac{1}{72}\pi i\right)+
 \frac{2}{3}\widehat{L_B}(\mu)\right]\,, \\
a^{(1)}_{C0}(s)&=&\frac{s^2}{32 \pi f^4}  
 \left[\frac{1}{16\pi^2}\left(\frac{83}{144}-\frac{4}{9}\log \left(\frac{s}{\mu ^2}\right)+\frac{1}{8}\pi i\right)+
 \frac{2}{3}\widehat{L_{C}}(\mu)\right] \,.
\end{eqnarray}
\end{widetext}
We defined the following combinations of Wilson coefficients:
\begin{eqnarray}
\widehat{L_A}(\mu)&=&\widehat{L_0}(\mu) + 68 \widehat{L_1}(\mu) + 36 \widehat{L_2}(\mu) + 17 \widehat{L_3}(\mu)\,, \nonumber \\
\widehat{L_B}(\mu)&=&2\widehat{L_0}(\mu) -4 \widehat{L_1}(\mu) + 2 \widehat{L_2}(\mu) - \widehat{L_3}(\mu)\,, \nonumber\\
\widehat{L_{C}}(\mu)&=&8\widehat{L_1}(\mu) +16 \widehat{L_2}(\mu) + 2 \widehat{L_3}(\mu) \,.
\end{eqnarray}
$\widehat{L_I}(\mu)$ are renormalized in the $\overline{MS}$ scheme and run according to the renormalization group equations, 
\begin{equation}
\widehat{L_I}(\mu)=\widehat{L_I}(\mu_0)+ \frac{k_I}{16\pi^2}\log \left(\frac{\mu}{\mu_0}\right)\,,
\label{eq:runwilson}
\end{equation}
with $k_I=-43/6$, $1/4$, $-4/3$ for $I=A$, $B$, $C$ respectively.
Of the 4 coefficients only 2 are independent for what concerns this process. In particular we have
\begin{equation}
\widehat{L_{C}}(\mu)=\frac{2}{7}\left(\widehat{L_A}(\mu)+10\widehat{L_B}(\mu)\right)\,.
\label{eq:LMSdep}
\end{equation}

\begin{figure}
\includegraphics[width=0.45\textwidth]{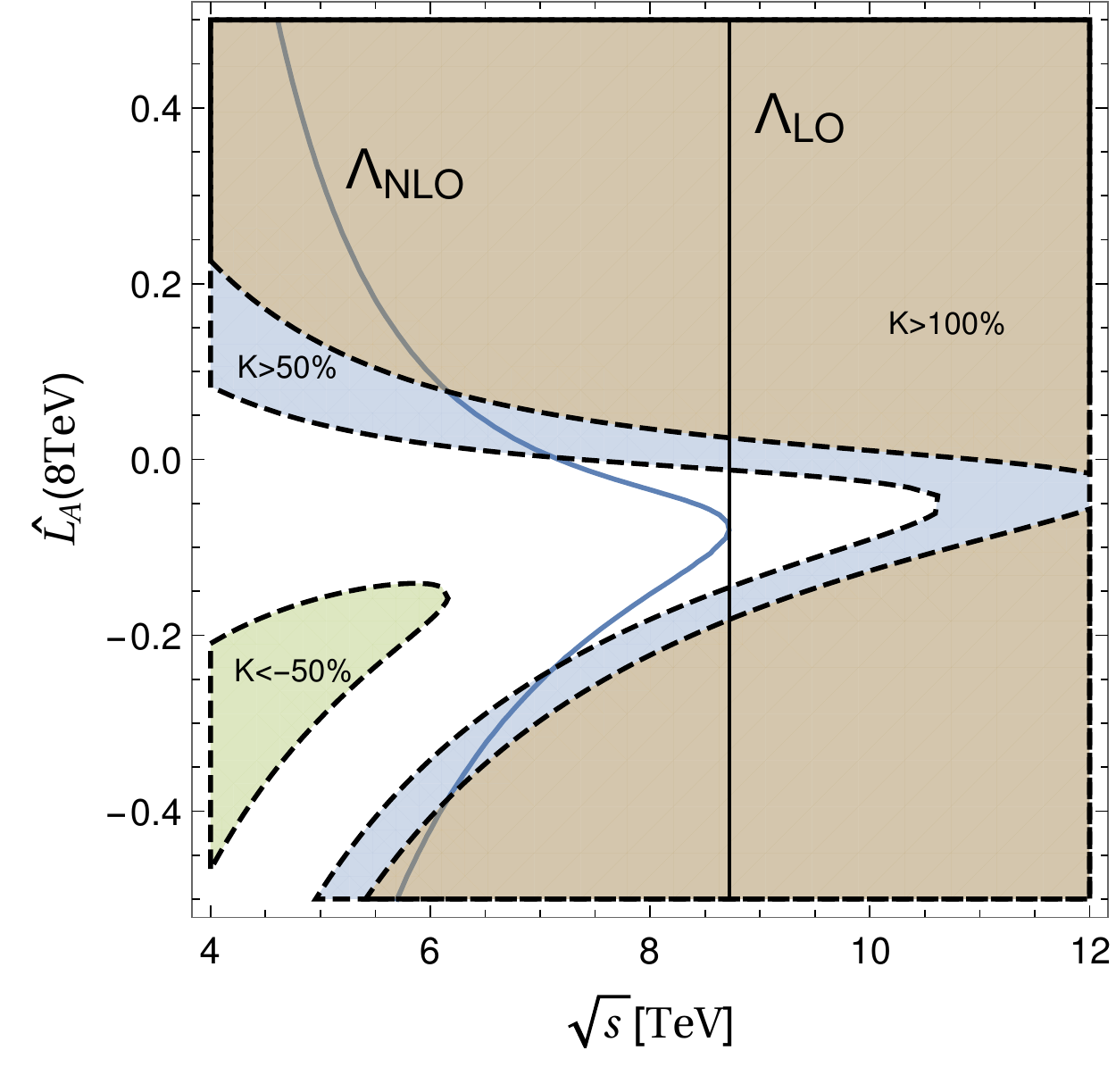}
\caption{Unitarity violation (\eq{eq:unitviolation}) scale of $a_{A0}(s)$ as a function of $\widehat{L_A}(8\TeV)$ for $\sin\theta=0.2$ at NLO (blue solid) and LO (black solid). Also shown the regions of lost of perturbativity $K\equiv|a^{NLO}(s)|/|a^{LO}(s)|-1>50\%$ (blue shaded area), $K>100\%$ (brown area) and $K<-50\%$ (green area). }
\label{fig:nlounit}
\end{figure}

The NLO and LO partial wave amplitudes can be defined by 
\begin{equation}
a^{NLO}(s)=a^{(0)}(s)+a^{(1)}(s),\quad a^{LO}(s)=a^{(0)}(s)\,.
\end{equation}
We show in \fig{fig:nlounit} the energy at which the $A0$-channel amplitude cross unitarity bounds, \emph{i.e.} $|a(s)|>1$, for $\sin\theta=0.2$~\footnote{The result for different $\theta$ is very similar, indeed if we choose the renormalization scale proportional to $\sqrt{s}$, $\mu\propto \sqrt{s}$ we find that the amplitudes depend only on the ratio $s/f^2$ apart from  logarithmic corrections from the running of the effective coefficients.}. 
At LO, the scale of unitarity violation, where $|a^{LO}(\Lambda_{LO})|=1$  is given in \eq{eq:lambdalo} and is independent of the Wilson coefficients, it is depicted as a vertical line in the figure. 
At NLO we show the scale of unitarity violation $|a^{NLO}(\Lambda_{NLO})|=1$ 
as a function of $\widehat{L_A}(8\TeV)$.

It can be noticed that NLO corrections always anticipate the violation of unitarity to lower scales, $\Lambda_{NLO}<\Lambda_{LO}$,  
and therefore $\Lambda_{LO}$ is an important physical scale. 
If the NLO corrections to $|a(s)|$ are positive, they will lead to a broad \emph{continuum} enhancement at least as strong as the LO amplitude or to resonance formation \emph{before} the scale of LO unitarity violation, \emph{i.e.}, the mass of the resonance obeys $M\lesssim \Lambda_{LO}$. For the scalar $A0$ channel this implies $M_\sigma\lesssim 1.7\TeV/\sin\theta$, which is more stringent  than lattice results on the scalar spectrum of $SU(2)$ gauge theory with 2 Dirac fermions which provide $M_\sigma=4.7(2.6)\TeV/\sin\theta$. 
If the NLO corrections are negative at $\sqrt{s}< \Lambda_{LO}$, there is a crossing point where $|a^{NLO}(s)|=|a^{LO}(s)|$ and the NLO amplitude must therefore have a faster growing behavior to fulfill $\Lambda_{NLO}<\Lambda_{LO}$ and trespass the LO amplitude before  $\Lambda_{LO}$. The NLO amplitude should therefore be controlled very likely by a more strongly bounded and narrower resonance. 
In either case, the LO amplitude enhancement can be regarded as the weakest and smoothest possible strong effect in GBS before unitarity violation.

We also show in shaded areas in the figure the regions where the $K$-factor  $K\equiv|a^{NLO}(s)|/|a^{LO}(s)|-1$ is $K>50\%$ (blue area), $K<-50\%$ (green area) and $K>100\%$ (brown area), where perturbativity is jeopardized. 
A correlation of lack of perturbativity and violation of unitarity is visible, as expected.

Inelastic channels do not change the upper limit on the scale of strong dynamics we just discussed, but they would anticipate the effects to lower energy scales, since they shrink the unitarity circle. 
The inelastic effect of $2\to 3$ NGB scattering at $\Lambda_{LO}$ is estimated to be $\sim 5\%$.


\subsection{Unitarization Model: the IAM method}
A  phenomenological approach to describe the physics beyond the perturbative regime in pion-pion scattering is given by Unitarization Models. They are based on formulas that force the amplitudes of GBS to satisfy the unitarity condition and maintain the low energy behavior. Unitarization models are intended to represent the approximate magnitude of these amplitudes beyond the perturbative regime and have been able in some cases to describe the first resonances of QCD.
In view of the great similarities between low energy QCD and the Electroweak physics, the ideas of unitarization models have been translated to a strong SB sector in many studies
\cite{Bagger:1993zf,Alboteanu:2008my,Butterworth:2002tt,Chanowitz:1995tn,Chanowitz:1996si,Hikasa:1991tw,Dobado:1999xb}. 
They are not complete quantum field theories and in particular they typically violate crossing symmetry, but despite those deficiencies, these models still carry out their phenomenological purpose of estimating the magnitude of strong $VV$ scattering cross sections much above the perturbative regime. 

One of the most widespread unitarization method used for $VV$ scattering is based on the  K-matrix, introduced already in the 40's~\cite{Schwinger:1948yk}. 
Besides violating crossing symmetry, the K-matrix unitarization procedure spoils the
singularity structure of the fixed-order amplitudes. 
Generalizations and improved versions of the method where analiticity is restored have been provided~\cite{Delgado:2013loa,Kilian:2014zja}.

In the $N/D$ protocol, unitarity is exactly restored with the extra quality of improved analytical properties. It is derived from dispersion relations~\cite{Chew:1960iv,Oller:1999me}. 

A special case of the $N/D$ method is the so-called Inverse Amplitude Method (IAM), which maintains the proper analytical structure of fixed order calculation with the correct branching cuts and without the need of extra parameters. It also produces very interesting phenomenological consequences in the context of strong vector boson scattering. It has been widely and successfully used in the description of low energy pion-pion scattering and has given remarkable results describing meson dynamics further beyond the perturbative regime, reproducing the first resonances in each isospin-spin channel up to 1.2 GeV. The method is derived from dispersion relation and can be regarded as a resumation of $s$-channel \emph{bubble} diagrams~\cite{Dobado:1992ha,Delgado:2015kxa}.
For certain values of the chiral coefficients, the unitarized amplitudes
present poles that can be interpreted as
dynamically generated resonances. 
The saturation of unitarity via resonances is indeed the expectation for typical strong dynamics~\cite{Gasser:1983yg,Ecker:1989yg}.

It is important to note that we use the unitarization method with caution, as a guidance of possible behavior of high energy amplitudes. We consider different values of effective coefficients and also non-resonant scenarios in order to encompass a complete range of viable strong effects. Therefore, we expect that a study using a different unitarization method (which at least respects the analytical structure of the amplitudes) should reproduce our results and conclusions. Indeed, in previous studies the improved K-Matrix method, the N/D and IAM have shown good agreement 
~\cite{Dobado:1999xb,Oller:1999me,Delgado:2015kxa}.
However, it has been advocated that the IAM is the correct one to treat the spin-1 channel~\cite{Delgado:2015kxa,Delgado:2017cls}.


These methods have been implemented and studied  in the context of the full $2\to 6$ matrix elements framework for strong VBS in Ref.~\cite{Ballestrero:2011pe}.

We will concentrate here on the IAM due to its good  analytical properties and the dynamical generation of resonances, which we aim to compare with the effective description of \sec{sec:effmodel}.
The IAM  defines the unitarized amplitude
\begin{equation}
a^{IAM}_{IJ}(s)=\frac{a^{(0)}_{IJ}(s)}{1-\frac{a^{(1)}_{IJ}(s)}{a^{(0)}_{IJ}(s)}}\,.
\end{equation}
For low energies this amplitude restores the chiral amplitudes while fully satisfying the unitarity condition.  
From the denominator of the IAM amplitudes a mass and a running width can be extracted
\begin{small}
\begin{eqnarray}
M_{A}^2&=&\frac{2f^2}{\frac{1}{16\pi^2}\left(\frac{29}{12}\right)+
 \frac{2}{3}\widehat{L_A}(M_A)},\quad  \Gamma_A= \frac{M_A^3}{16\pi f^2},\nonumber\\
 \label{eq:massI}
M_{B}^2&=&\frac{(f^2/6)}{\frac{1}{16\pi^2}\left(-\frac{35}{432}\right)+
 \frac{2}{3}\widehat{L_B}(M_B)},\quad  \Gamma_B= \frac{M_B^3}{192\pi f^2}, \nonumber\\
M_{C}^2&=&\frac{-(f^2/2)}{\frac{1}{16\pi^2}\left(\frac{83}{144}\right)+
 \frac{2}{3}\widehat{L_{C}}(M_{C})} ,\quad  \Gamma_{C}= \frac{M_{C}^3}{64f^2}\,.
 \label{eq:massMS}
\end{eqnarray}
\end{small}
The amplitudes can then be written in a particularly simple form by choosing a dynamical renormalization scale $\mu=\sqrt{s}$,
\begin{small}
\begin{equation}
a^{IAM}_{IJ}(s)=\frac{-\Gamma_I/M_I} 
{s-M_I^2+i \frac{\Gamma_I}{M_I}s + 32\pi s \frac{\Gamma_I}{M_I}\frac{k_I}{16\pi^2} \log\left(\frac{\sqrt{s}}{M_I}\right)}
\end{equation}
\end{small}
with $k_I$ given in \eq{eq:runwilson}.

\begin{figure}
\includegraphics[width=0.45\textwidth]{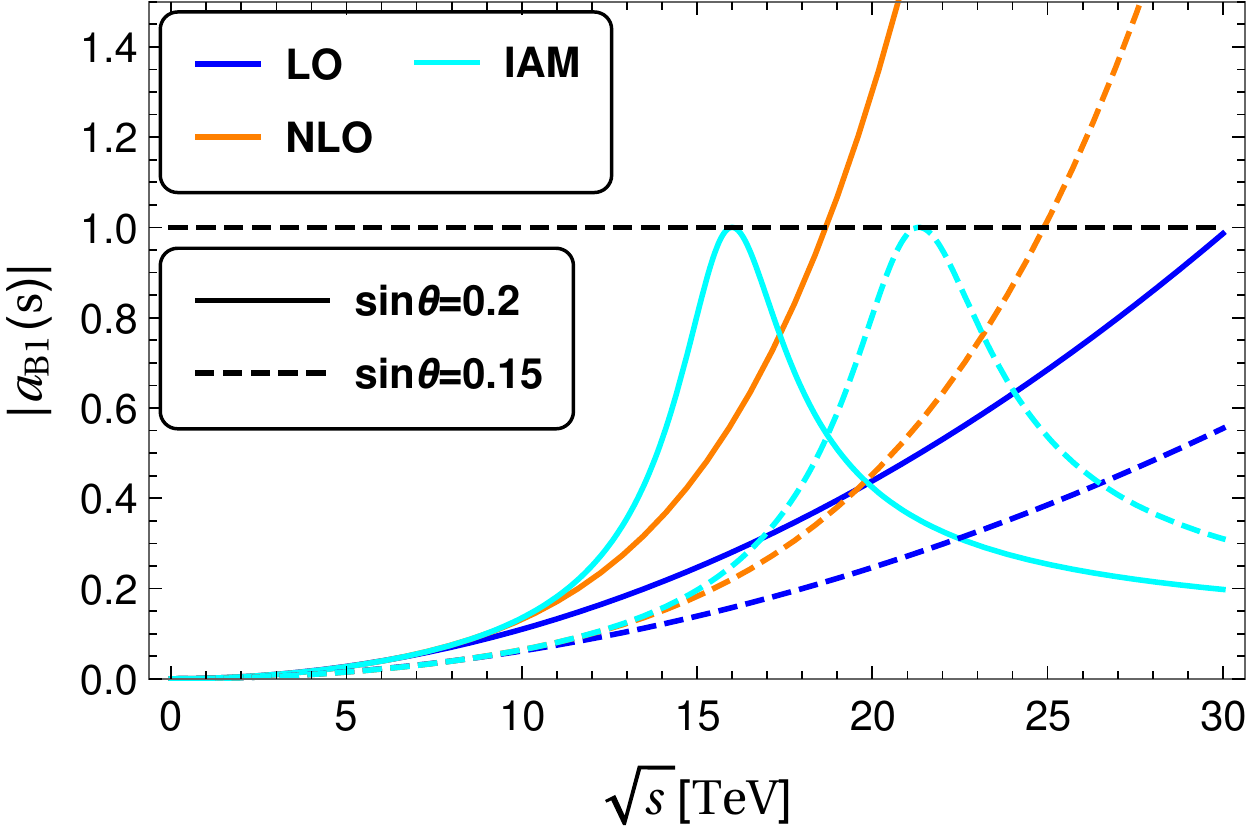}
\caption{Absolute value of partial wave amplitude $a^{IAM}_{B1}(s)$ shown together with $a^{NLO}_{B1}(s)$ and $a^{LO}_{B1}(s)$ for $\sin\theta=0.2$ and 0.15.}
\label{fig:aS1IAM}
\end{figure}

As a specific example and benchmark scenario we will now make use of lattice results  $M_V\equiv M_B = 3.2(5)\TeV/\sin\theta$.
 The corresponding effective coefficient can be extracted from \eq{eq:massMS}, $L_B(M_V)=2.225\times 10^{-3}$, and it is independent of $\theta$.
The $J=1$ partial wave amplitude for this scenario is shown for $\sin\theta=0.2 (0.15)$ in \fig{fig:aS1IAM}. We use renormalization scale $\mu=\sqrt{s}$.

\begin{figure}
\includegraphics[width=0.45\textwidth]{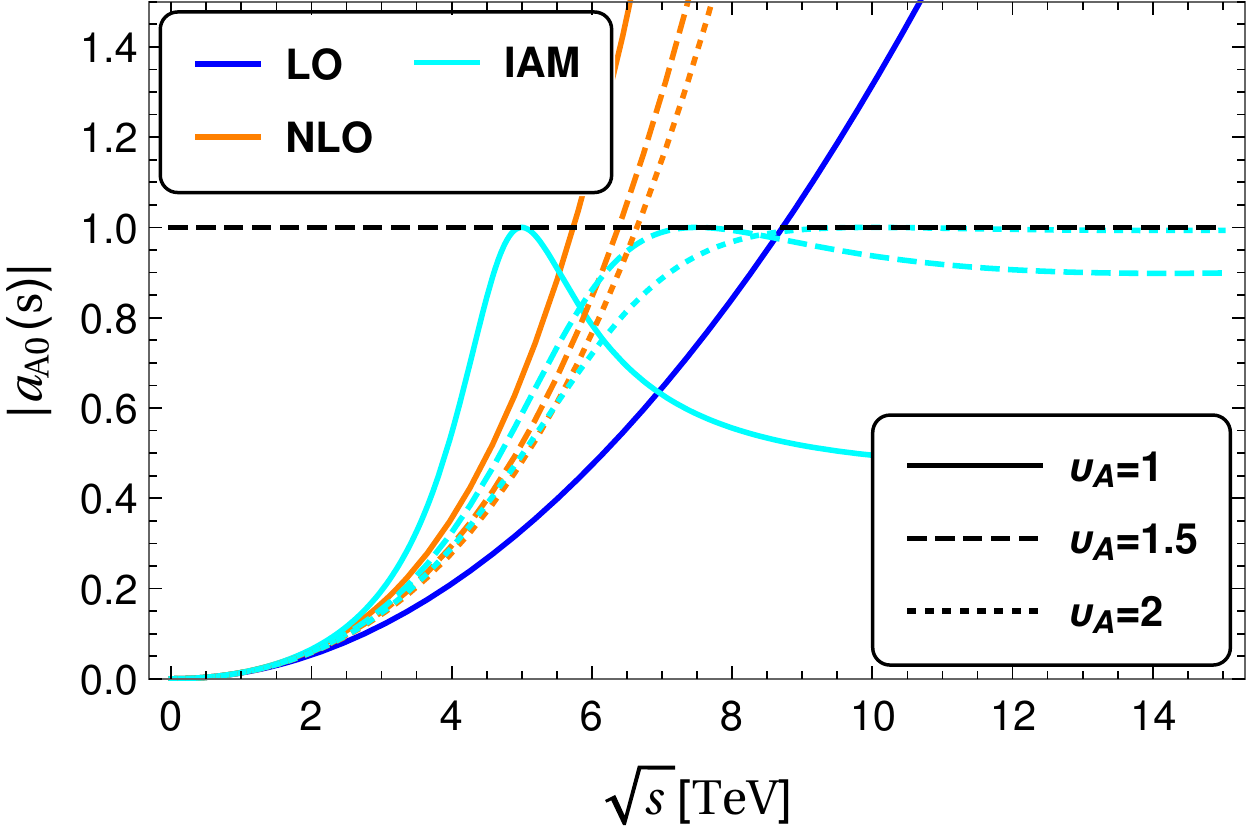}
\caption{Absolute value of partial wave amplitude $a^{IAM}_{A0}(s)$ shown together with $a^{NLO}_{A0}(s)$ and $a^{LO}_{A0}(s)$.}
\label{fig:aI0IAM}
\end{figure}

For the scalar channel the lattice result $M_\sigma\equiv M_A=4.7(2.6)\TeV/\sin\theta$ has very large uncertainty.
The mass of an eventual resonance is also proportional to the scale $f$, thus we define 
the parameters 
\begin{equation}
\upsilon_I \equiv\frac{M_I\sin\theta}{\TeV},\quad I=A,B,C\,.
\end{equation}
The effective coefficient for channel $A$ is 
$L_A(M_A)=-0.0229556 + 0.181548/\upsilon_A^2$.
The corresponding unitarized amplitude is shown in \fig{fig:aI0IAM} for different values of $\upsilon_A$. For large values of $\upsilon_A\gtrsim 1.5$ a broad enhancement takes the place of the typical Breit-Wigner peak of a  resonance.  

\begin{figure}
\includegraphics[width=0.45\textwidth]{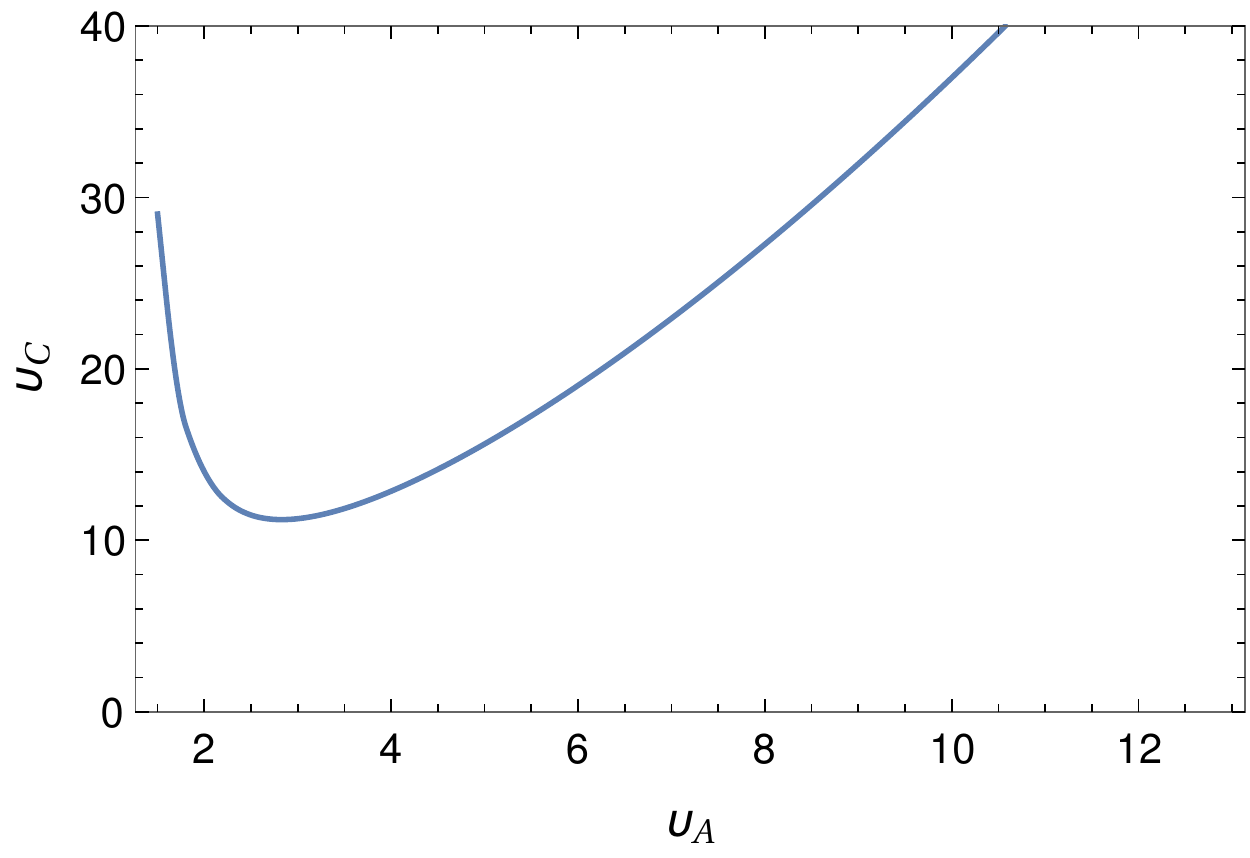}
\caption{Value of $\upsilon_{C}$ as a function of $\upsilon_{A}$ for fixed value of $\upsilon_B=3.2$.}
\label{fig:aMS0IAM}
\end{figure}

We now look at unitarization of the $C$ channel. As mentioned before, the effective coefficients are linearly dependent according to \eq{eq:LMSdep}; therefore, if we choose to fix lattice inspired $\upsilon_B=3.2$, we find the relation among $\upsilon_{C}$ and $\upsilon_A$, shown in \fig{fig:aMS0IAM}.
We conclude that this eventual resonance must be at higher scales. 

\subsection{Vector Resonances}
\label{sec:vector}

In this and the following sections, we will use the chiral Lagrangians described in \sec{sec:effmodel} including vector and scalar states to estimate their parameters in the light of unitarity considerations just explored. Let us start with the vector case.  

\begin{figure}
\includegraphics[width=0.45\textwidth,height=0.3\textwidth]{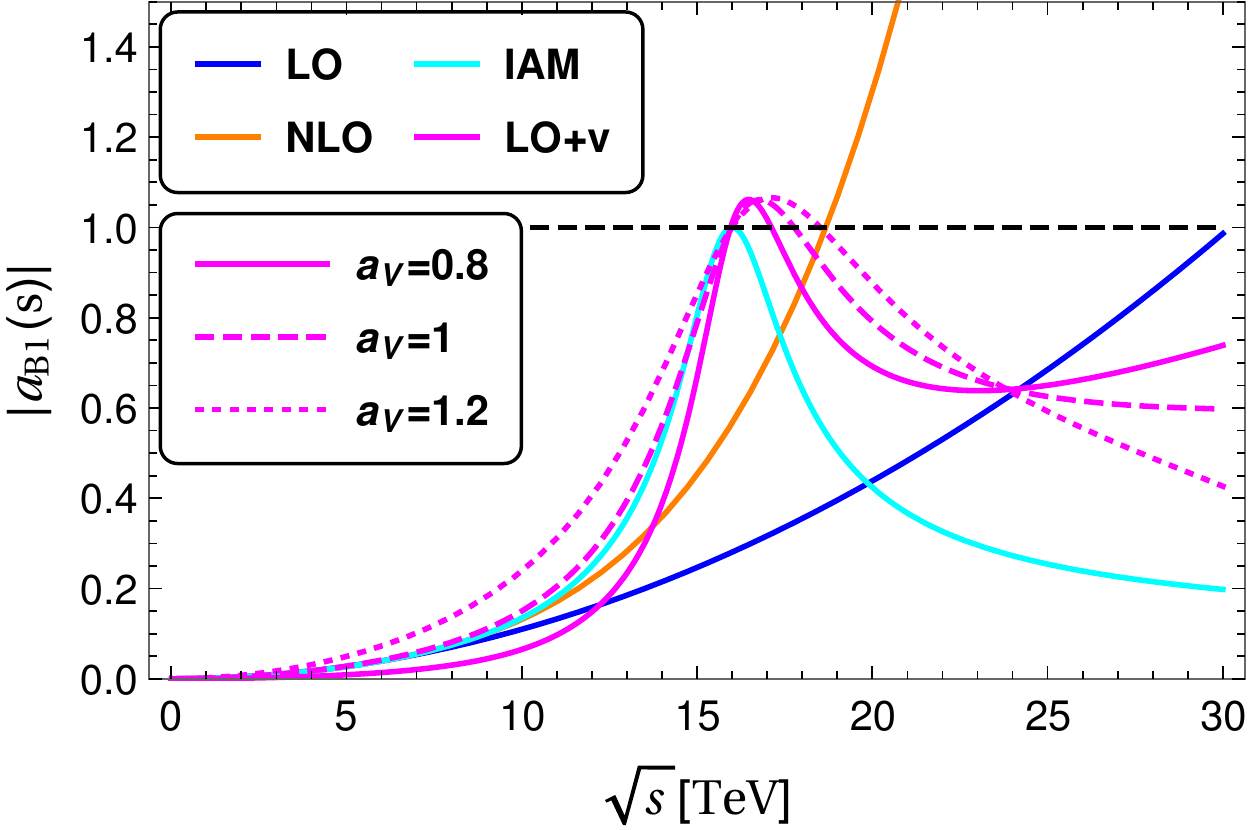} 
\caption{Absolute value of partial wave amplitude $a^{0}_{B1}(s)+a^{v}_{B1}(s)$ together with LO, NLO and IAM equivalents, for three values of $a_V=0.8$, 1, 1.2  and $\sin\theta=0.2$.}
\label{fig:aS1v}
\end{figure}

At tree-level the projections can all be computed from the single master amplitude $\mathcal{A}(s,t,u)$ of the process $\pi^+\pi^-\to \pi^0\pi^0$. 
The vector states contribute with trilinear couplings (\eq{eq:vtril}) to the NGBs and also by modifying the  quartic coupling of NGBs to recover the correct LET behavior, giving
\begin{equation}
\mathcal{A}(s,t,u)= - g_V^2\left(\frac{s-u}{t-M_V^2}+\frac{s-t}{u-M_V^2} +\frac{3s}{M_V^2} \right)\,.
\end{equation}
The projections are given by
\begin{small}
\begin{eqnarray}
\mathcal{A}_A(s,t,u)&=& 5\mca(s,t,u)+\mca(t,s,u)+\mca(u,t,s)\,,\nonumber\\
\mathcal{A}_B(s,t,u)&=& \mca(t,s,u)-\mca(u,s,t)\,,\nonumber\\
\mathcal{A}_{C}(s,t,u)&=& \mca(t,s,u)+\mca(u,s,t)\,. 
\end{eqnarray}
\end{small}
Further expanding in partial waves we get 
\begin{widetext}
\begin{eqnarray}
a^v_{A0}(s)&=&-\frac{g_V^2}{8\pi}\left[ (2  + 3\frac{s}{M_V^2}) 
             - 2(\frac{M_V^2}{s}+2)\log(1+\frac{s}{M_V^2}) \right]\,,\\
a^v_{B1}(s)&=&\frac{g_V^2}{32\pi}\left[ \frac{s}{3(s-M_V^2)}-\frac{s}{2M_V^2} 
             - (\frac{M_V^2}{s}+2)\left(2 - (2\frac{M_V^2}{s}+1)\log(1+\frac{s}{M_V^2})\right) \right]\,.
\end{eqnarray}
\end{widetext}
The $J=1$ amplitude is shown in \fig{fig:aS1v} for $\sin\theta=0.2$ and lattice inspired value of mass $M_V=3.2\TeV/\sin\theta$. We show 3 different values of the vector coupling $a_V=0.8$, 1, 1.2.
We can see that $a_V$ must be close to 1 to better describe  the dynamical inspired IAM amplitude. Moreover, the departure from $a_V=1$ creates large deviations from the LO amplitude at low energy. 
We thus take $a_V=1$ as a \emph{natural} value.
The total width of the decay into the NGBs is given by $\Gamma_V=\frac{g_V^2}{48\pi}M_V$. The choice 
$a_V=1$ reproduces the total width provided by IAM method, \eq{eq:massMS}.

A remark about the EW and photon exchange follows. It is well known that low-mass boson exchange leads to large logarithmic enhancements due to the so-called ``exceptional" phase-space regions, \emph{e.g.}, the terms $\log(1+\frac{s}{M_V^2})$ when $t\to 0$ . These large logarithms need usually to be resummed for improved perturbative calculations. Nevertheless, this EW physics is not relevant for the present analysis, as customary~\cite{Cornwall:1974km,Lee:1977eg}. 

\subsection{Scalar isosinglet $\sigma$}
\label{sec:sigma}

The $\sigma$ contribution to the master amplitude is given by
\begin{equation}
\mathcal{A}(s,t,u)=-g_\sigma^2\frac{s}{f^2}\frac{s}{s-M_\sigma^2}\,,
\end{equation}
with $g_\sigma=\kappa'/2$.

The total width of $\sigma$ into NGBs is given by
$\Gamma_\sigma=5\frac{g_\sigma^2 m_\sigma^3}{32\pi f^2 }$. Requiring a width similar to IAM leads to
$g_\sigma\sim 0.63$. 

\begin{figure*}
\includegraphics[width=0.45\textwidth]{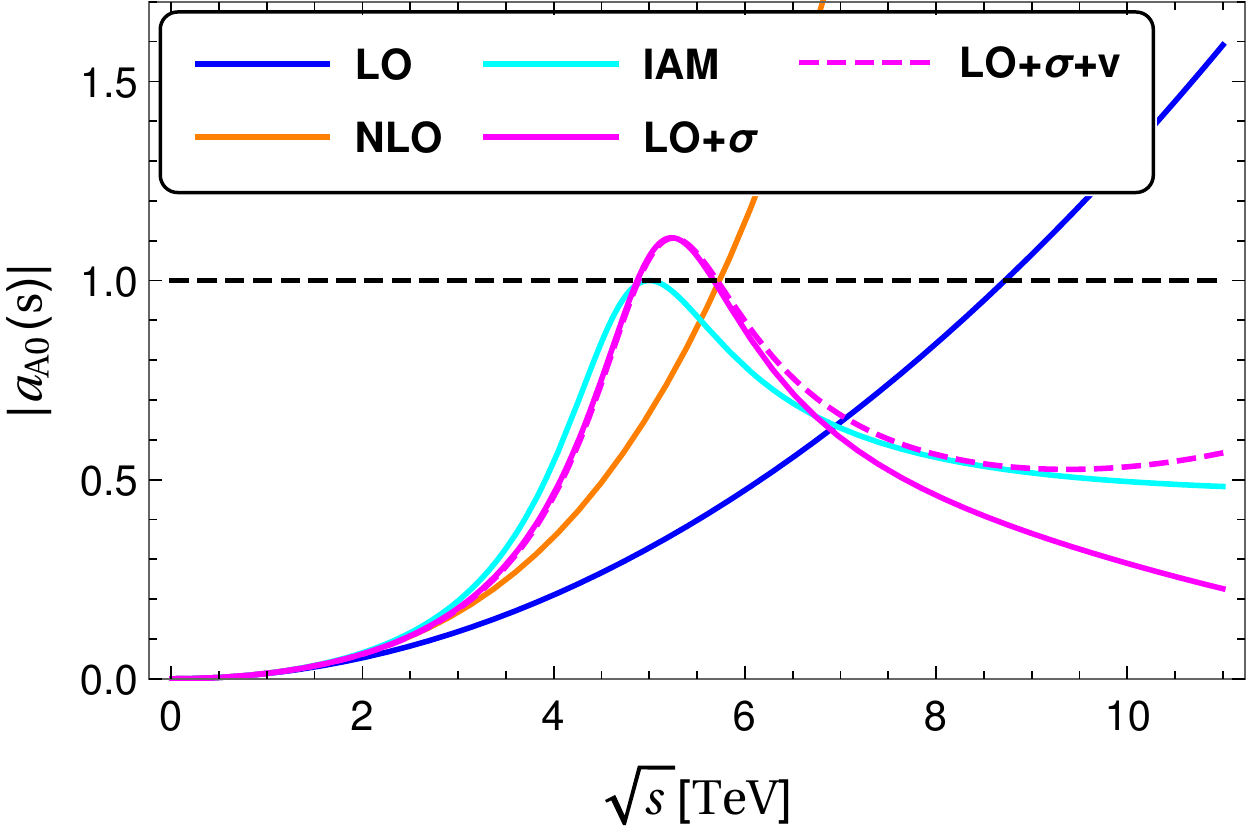}
\includegraphics[width=0.45\textwidth]{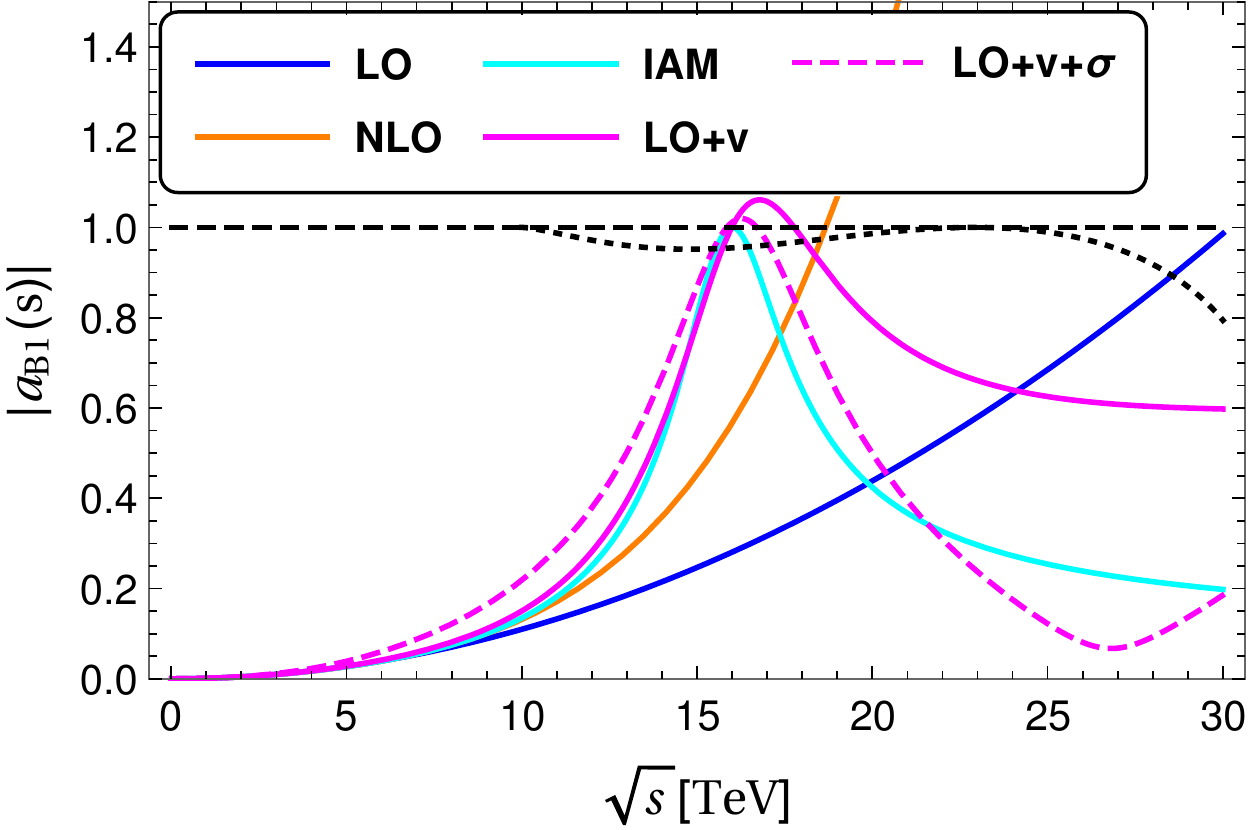}
\caption{\emph{Left} panel: absolute value of partial wave amplitudes $a^{0}_{A0}(s)+a^{\sigma}_{A0}(s)+a^{v}_{A0}(s)$ together with LO, NLO and IAM equivalents. \emph{Right} panel: equivalent amplitudes for $B1$ channel, $a^{0}_{B1}(s)+a^{\sigma}_{B1}(s)+a^{v}_{B1}(s)$. Parameters are $\upsilon_A = 1$, $\sin\theta=0.2$, $g_\sigma=0.63$. The effect of inelastic channels in the unitarity bound is depicted in dotted black curve and explained in the text.}
\label{fig:aI0sigma}
\end{figure*}

We show on the left-hand panel of \fig{fig:aI0sigma} the $a_{A0}(s)$ amplitudes, including the $\sigma$ contribution for $g_\sigma=0.63$ and $\upsilon_A = 1$. 
We show also the contribution from the $v$ state with $a_V=1$.
On the right-hand panel, we show the equivalent $a_{B1}(s)$ amplitudes.
The effect of the inelastic channel $\pi\pi\to \sigma\sigma$ in the unitarity bound is depicted in the dotted black line in the figure. The corresponding amplitude was computed assuming $\kappa''=1.5$ and a $\sigma^3$ trilinear coupling  $\lambda_3=1$ ($g_\sigma=0.63$ and $\upsilon_A = 1$ are kept unchanged), and its contribution to $B1$ channel estimated. The new unitarity bound is taken to be $|a(s)|<1/2+\sqrt{1/4-|a^{2\sigma}(s)|^2\sqrt{1-4M_\sigma^2/s}}$, according to
\eq{eq:pwinel}. It can be noticed that the effect is small for this value of $\kappa''$.
We also note that interactions between $\sigma$ and vector states are viable but were not considered here for simplicity.

\begin{figure}
\includegraphics[width=0.45\textwidth]{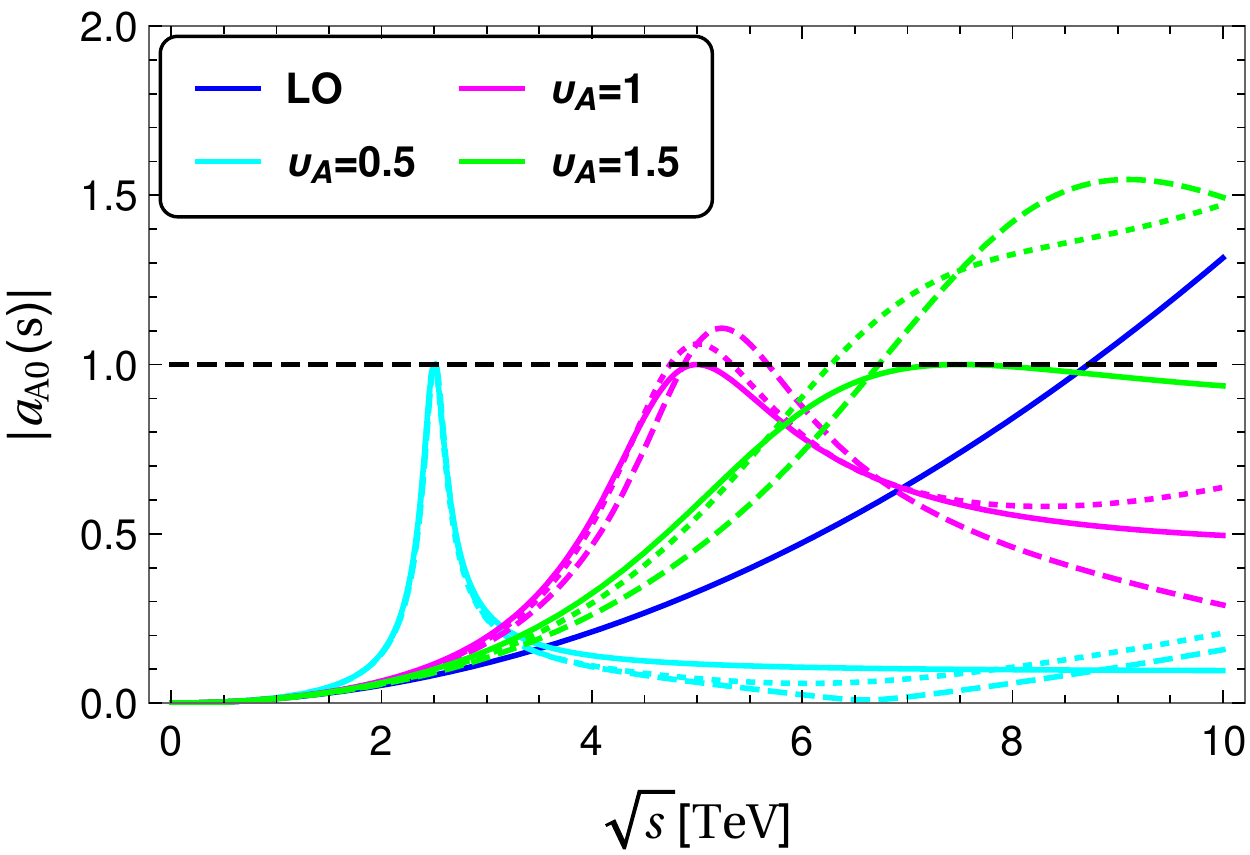} 
\caption{Absolute value of partial wave amplitudes $a^{0}_{A0}(s)+a^{\sigma}_{A0}(s)$ for $\upsilon_A = 0.5$, 1, 1.5, using the IAM unitarization model (solid curve), a fixed width $\sigma$ resonance (dashed) and a running width (dotted).}
\label{fig:aI0sigmaRun}
\end{figure}

For larger values of $\upsilon_A\gtrsim 1$ the growing behavior of the LO piece renders difficult for a resonance to unitarize the amplitude.
This fact is illustrated in \fig{fig:aI0sigmaRun} where we show the $a_{A0}(s)$ amplitudes for 3 values of $\upsilon_A=0.5$, 1, 1.5, using the IAM unitarization model (solid curve), the fixed width $\sigma$ resonance (dashed) or a running width, $\Gamma_{fix}\to \frac{\Gamma_{run}}{M}s$, (dotted)~\footnote{
The resummation of the self-energy diagrams lead to momenta dependent widths, or \emph{running widths}, which are typically important for heavy and broad resonances. An \emph{ad-hoc} incorporation of such running width is however not usually recommendable due to large extra mis-cancellations which  worsen unitarity problems at higher energies, and can be cured with running width gauge invariant method as in Ref.\cite{Franzosi:2012nk}.}. 
It can be seen that values of $M_\sigma$ too close or larger than unitarity violation scale, $M_\sigma\sim \Lambda_{LO}$, prevent any meaningful use of resonant propagation and a broad \emph{continuum} appears instead. Moreover, close to the peak the running width approach slightly ameliorates the lineshape description. 

\begin{figure}
\includegraphics[width=0.45\textwidth]{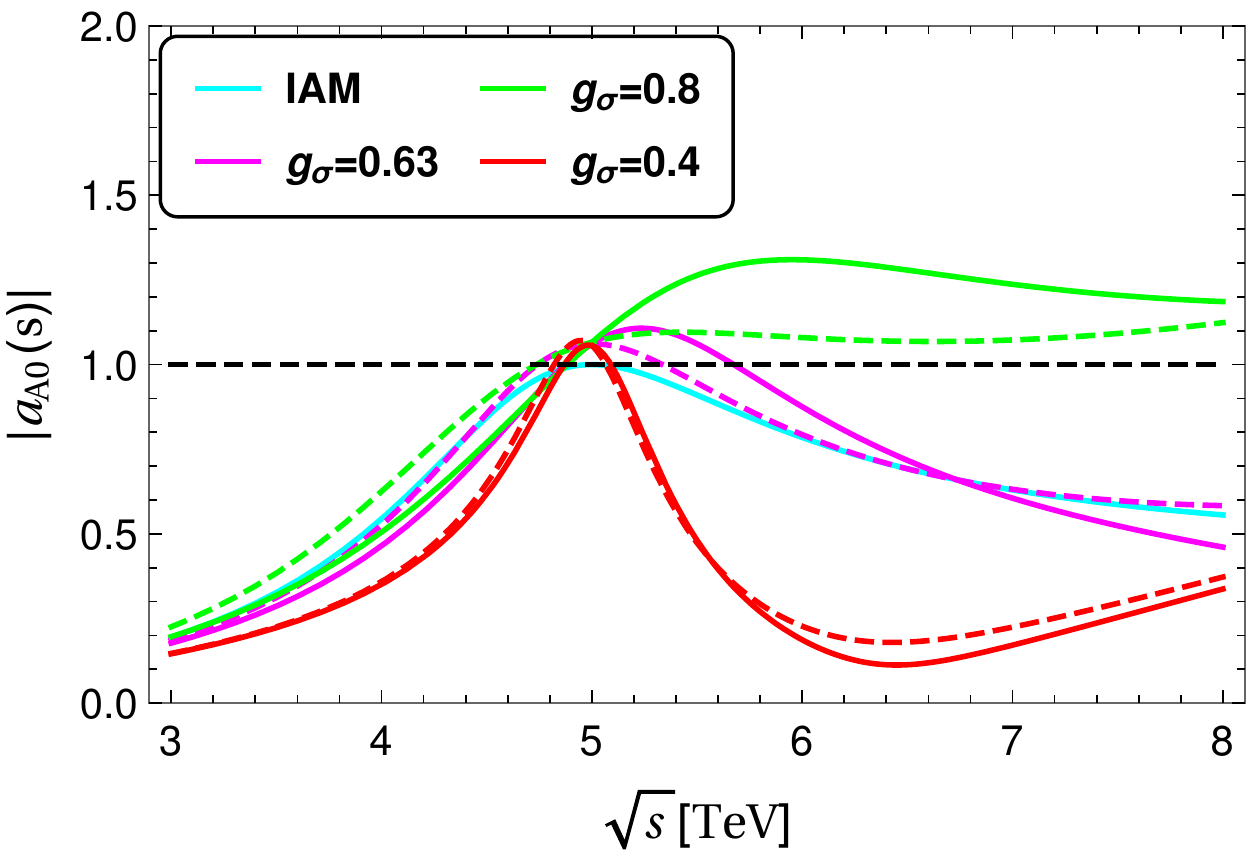} 
\caption{Absolute value of partial wave amplitudes $a^{0}_{A0}(s)+a^{\sigma}_{A0}(s)$ for $g_\sigma = 0.4$, 0.63, 0.8, using a fixed width (solid curve) and a running width (dashed) compared to the IAM (cyan).}
\label{fig:aI0sigmaRungs}
\end{figure}

Similarly, large couplings can also jeopardize the resonant description and violate unitarity. Extra contributions to the width can dump down and unitarize the amplitude, but nevertheless not helping in the description of the lineshape. 
In \fig{fig:aI0sigmaRungs} we show the $a_{A0}(s)$ amplitudes for 3 values of $g_\sigma$ using  the IAM unitarization model (cyan), a fixed width (solid) and a running width (dashed).

\section{Experimental signatures at Future Colliders}
\label{sec:experimental}


We found in previous sections that the dynamically inspired parameters are $a_V\sim 1$ and $M_V\sim 3.2\TeV/\sin\theta$ in the vector sector and $g_\sigma=0.63$ and $M_\sigma\lesssim 1.2\TeV/\sin\theta$ in the scalar sector. Alternatively, LET behavior gives a meaningful benchmark scenario for non-resonant \emph{continuum} (below unitarity violation). In all cases, $\sin\theta<0.2$. In this section we study these scenarios in realistic  observables at hadron colliders.

Composite vector states can have large mixing with the SM weak bosons, which generates minimal coupling to fermions\footnote{We assume there is no direct coupling to fermions, even though this is a logical possibility.}, and consequently complementary production modes, either via DY or via Vector Boson Fusion (VBF), as well as complementary decay modes into fermions or bosons. We will discuss the vector phenomenology in \sec{sec:vectorpheno}.
Similarly, the $\sigma$ scalar resonance mixes with the Higgs boson and generates minimal couplings to SM fermions proportional to their masses, which would lead to its production through gluon fusion via a loop of top-quarks. However, this mixing should be small and the dominant channel has to be VBF production with decay to weak bosons.
This signature falls in the same class of process of strong VBS, $VV\to VV$.
Due to the intrinsic high compositeness scale  of CH models for $\sin\theta\lesssim 0.2$, these typical strong effects will be observable more likely at a future 100 TeV machine than at the LHC. 

In proton-proton collisions, VBS is embedded in more complicated processes where a quark in each proton emits a gauge boson, $V$. These scatter among themselves and produce two $V$s along with the 2 extra remnant jets in the forward-backward region of the detector. The $V$s subsequently decay into jets and/or leptons. 
This process has been scrutinized since a long time
\cite{Duncan:1985vj,Dicus:1986jg,Cahn:1986zv,Barger:1987du,Kleiss:1987cj,Barger:1988mr,Barger:1990py,Baur:1990xe,
Dicus:1990fz,Barger:1991ar,Dicus:1991im,Barger:1994zq,Bagger:1995mk,Rainwater:1999sd,Accomando:2005hz, Ballestrero:2008gf,Ballestrero:2010vp,Ballestrero:2012xa}
with an increasing degree of sophistication, in particular in the context of CH models~\cite{Ballestrero:2009vw,Contino:2011np}  and for Walking Technicolor with the Higgs identified as the first scalar excitation~\cite{Franzosi:2012ih}, and more recently at a 100 TeV collider~\cite{Szleper:2014xxa,Mangano:2016jyj,Jager:2017owh}.


The goal we will pursue in \sec{sec:vbfpheno} is to assess the possibility to distinguish the CH scenario from the SM predictions looking at the high energy region of $M(VV)$. In the CH scenario an overall excess or a resonance are expected. 
We consider only the simplest and cleanest VBS channel where 2 $Z$ decay into leptons, $pp\to jj ZZ\rightarrow jj4\ell$. 
The only relevant backgrounds are SM electroweak  $ZZjj$ and QCD $ZZ$+jets production. 
Other VBS channels, $WW$, $WZ$ and other decay channels will definitely improve the discriminant power here presented~\cite{Jager:2017owh}.

\subsection{Vector Phenomenology}
\label{sec:vectorpheno}


To get cross sections and branching ratios (BR) for the composite vectors we make use of the full model presented in Ref.\cite{Franzosi:2016aoo} and briefly described in \app{app:ccwz}. It was implemented in the UFO format~\cite{Degrande:2011ua} via the \FeynRules package~\cite{Alloul:2013bka} and is available in the HEPMDB\footnote{http://hepmdb.soton.ac.uk/hepmdb:0416.0200}. We use the  PDF set NNPDF 2.3 at LO~\cite{Ball:2012cx}\footnote{Only the first two families of quarks are included, even though the third family is known to be important for a center of mass energy of 100 TeV~\cite{Han:2014nja}.}. We use \Madgraph~\cite{Alwall:2014hca} to compute the cross sections for both DY and VBF productions. For the calculation of VBF cross sections we have selected the minimum set of gauge invariant diagrams in $pp\to VVjj$ which contains the VBF topology and applied a minimum transverse energy on the jets, $p_T(j)>20\GeV$, to avoid singularities.

The heavy masses of these states $M_V\gtrsim 16\TeV$ (since $\sin\theta\lesssim 0.2$) have to be probed at higher energies than those available at the LHC. A 100 TeV machine like the FCC is the natural candidate. 
The limits on production cross section times branching ratio ($\sigma\times BR$)  of general vectorial resonances $\rho$ at the FCC have been  derived in Ref.\cite{Thamm:2015zwa}. This study is based on the exclusion sensitivities of two \LHC analyses \cite{Khachatryan:2014sta,Khachatryan:2014xja} and on the scaling of cross sections due to the  evolution of the parton luminosities.
The limits are provided as a function of the resonance mass, $M_\rho$, for two different decay channels: $\rho \rightarrow \ell^+\ell^-$ and $\rho \rightarrow WZ$, and two integrated luminosities $L = 1 \, ,\, 10 \, \text{ab}^{-1}$.
In \tab{tab:ex_limits} we show the exclusion limits at 95\%CL on $\sigma \times \text{BR}$  for $\sin\theta=0.2$, corresponding to $M_\rho\sim 16\TeV$ (apart from mixing effects), and $\sin\theta=0.15$ with 
$M_\rho\sim 21.3\TeV$.

\begin{table}[h]
\begin{tabular}{|c|c|c|c|}
\hline
$L [\text{ab}^{-1}]$ & decay & $M_\rho= 16\TeV$ & $M_\rho= 21.3\TeV$\\ 
\hline
1 & $\ell^+\ell^-$ & $2.28\times10^{-6}$ pb & $3.7\times10^{-6}$ pb\\
10 & $\ell^+\ell^-$ & $4.01\times 10^{-7}$ pb & $7.49\times10^{-7}$ pb\\
\hline
1 & $WZ$ & $4.0\times 10^{-4}$ pb & $3.78\times10^{-4}$ pb\\
10 & $WZ$ & $3.73\times10^{-5}$ pb & $5.41\times10^{-5}$ pb\\
\hline
\end{tabular}
\caption{Exclusion limits at 95\%CL on the $\sigma \times$ BR  production process $pp\rightarrow \rho$, in two different decay modes, $\rho\rightarrow \ell^+\ell^-$ and $\rho \rightarrow WZ$. Values for two different luminosities, $1$ and $10 \, \text{ab}^{-1}$ and two different masses, $M_\rho=16\TeV$ (21.3 TeV) are extracted from \cite{Thamm:2015zwa}.}
\label{tab:ex_limits}
\end{table}

From the vast spectrum of 15 vector states, the iso-triplet $V^{0,\pm}$ will be the most massively produced. The second triplet $S^{0,\pm}$ is only a bit heavier, near degenerate with $V^{0,\pm}$, but has lower cross section into fermions and weak bosons. It could dominate in the Higgs decay channels, which were not considered in Ref.\cite{Thamm:2015zwa}. The $A^{0,\pm}$ states will also be produced in a proton-proton collision, but they are heavier and will be more difficult to observe. Other states do not mix with SM particles and are much harder to be produced. Therefore, it is safe to assume the first observed peak will come from the $V^{0,\pm}$ states and we will neglect the other contributions.

Once $\sin\theta=0.2,\,0.15$, $M_V=3.2\TeV/\sin\theta$ and $M_{\bm{\mathcal{A}}}=3.5\TeV/\sin\theta$ are fixed, we are left with 2 extra free parameters: $\gt$ and $r$. For $r=1$ the decay into fermions dominates. Once $r$ departs from 1, the diboson decay channel becomes more important and rapidly overcomes the fermion channel. 

\begin{figure*}
\begin{tabular}{c c}
\includegraphics[width=0.5\textwidth,height=0.4\textwidth]{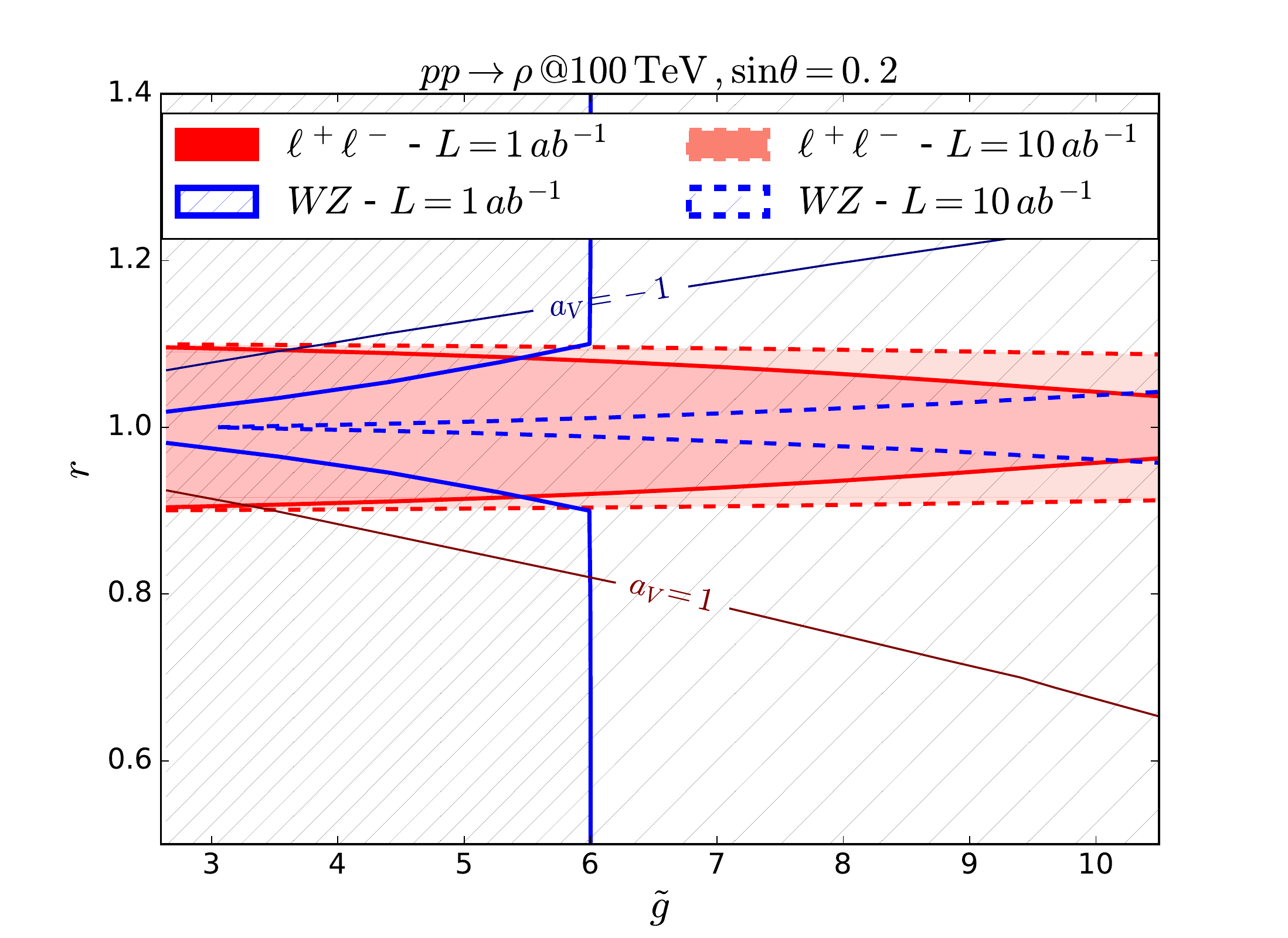} &
\includegraphics[width=0.5\textwidth,height=0.4\textwidth]{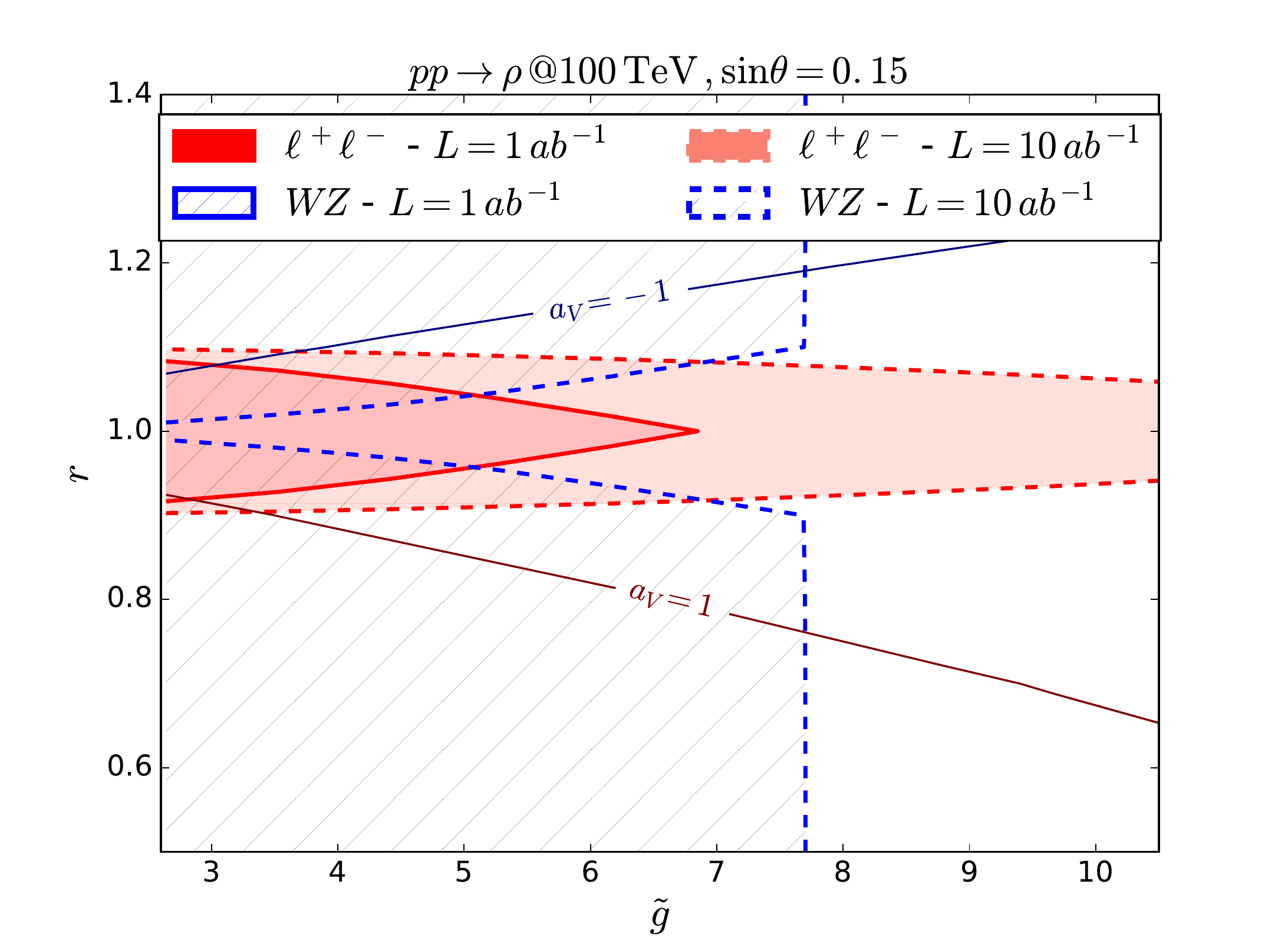} \\
(a) & (b) 
\end{tabular}
\caption{95\%CL exclusion regions in $(\gt,r)$-plane for $\sin\theta=0.2$ (\emph{left}) and $\sin \theta=0.15$ (\emph{right})  in the $\rho\to WZ$ (blue contour, hashed) and $\rho\to \ell^+\ell^-$ (red, shaded) channels.
  $L = 1\, \text{ab}^{-1}$ (solid contour) and $ L=10\,\text{ab}^{-1}$. }
  \label{fig:exclusion_limit_1}
\end{figure*}

In \fig{fig:exclusion_limit_1} we can see the  excluded region at $95 \%$ of confidence level in the plane $(\gt,r)$ for $\sin\theta=0.2$ (\emph{left} panel) and $\sin\theta=0.15$  (\emph{right} panel). 
The full parameter space for $\theta=0.2$ can be excluded with a luminosity $L = 10 \, \text{ab}^{-1}$ (dashed line).  For $\theta=0.15$ there is a region $\gt \gtrsim 8$ and $ |r-1| \gtrsim 0.1$ which will not be excluded with $10\iab$.

Lines of dynamically inspired $|a_V| = 1$ are also depicted in the plots.

\subsection{Strong VBS in $pp\to jjZZ\to jj4\ell$}
\label{sec:vbfpheno}

To model the non-resonant excess and the $\sigma$ resonance we have implemented the Lagrangian in \eq{eq:sigma} in the UFO format via the \FeynRules package. We consider the following benchmark scenarios:

{\bf LET non-resonant enhancement:} this is the Lagrangian in \eq{eq:sigma} without $\sigma$. It is the simplest and most conservative effect of strong VBS in CH models and is a general feature not specific to the SU(4)/Sp(4) realization. The observation of this excess gives an indirect probe of the Higgs coupling to weak bosons \cite{Szleper:2014xxa}. At the \LHC, the measurement of $hZZ$ coupling can reach  3\% accuracy in the most optimistic case or 5\%  in a more realistic scenario (at 1 standard deviation)~\cite{Mariotti:2016owy}. These deviations correspond to $\sin\theta\sim 0.24$ and $\sim 0.31$ respectively. We will show that even with only the $ZZ\to 4\ell$ channel we may exclude $\sin\theta= 0.2$ with good probability.  We consider also $\sin\theta= 0.15,\, 0.1$. 
We note that the energies beyond LO unitarity violation have negligible contribution for our analysis.

{\bf Scalar $\sigma$ resonance:}  we summarize in \tab{tab:sigma_benchmark} the benchmark scenarios we have considered.
The first 4 scenarios in the table will be analyzed for a 100 TeV machine and the last one is an optimistic case to be analyzed at LHC energies.

\begin{table}[h]
\begin{tabular}{c c c c c c}
\hline
$\sin\theta$ & $\upsilon_A$ & $M_{\sigma}$[TeV]& $\Gamma_\sigma$[TeV]& $g_\sigma$ & collider\\
\hline
 0.2 		& 1.2 & 6	& 2.81 & 0.63 & FCC \\
 0.15   		& 0.9 & 6	& 1.58 & 0.63 & FCC\\
 0.1 		& 0.6 & 6	& 0.7  & 0.63 &FCC\\
 0.1 		& 0.8 & 8	& 2.69 & 0.8 & FCC\\
 0.2 		& 0.8 & 4   & 1.34 & 0.8 & LHC \\
\hline
\end{tabular}
\caption{Parameters of benchmark scenarios for the CH model with $\sigma$ resonance.}
\label{tab:sigma_benchmark}
\end{table}

Events for the process $pp\to jjZZ\rightarrow e^+ e^- \mu^+ \mu^- jj$ have been simulated at LO with the multi-purpose generator \Sherpa \cite{Gleisberg:2008ta}. We imported the UFO model through the BSM module \cite{Hoche:2014kca} available for the \Comix matrix element generator \cite{Gleisberg:2008fv}.  All the samples generated have LO accuracy, and are showered through the  \CSS module, the Catani-Seymour dipole based shower \cite{Schumann:2007mg}. 
We have used dynamical factorization and renormalization scales $\mu^2_F = \mu^2_R = \left( p^{\mu}_{Z_1} + p^{\mu}_{Z_2}\right)^2$. The NNLO CT14 PDF set ~\cite{Dulat:2015mca} in the 4 flavor scheme has been employed~\footnote{Here again the 3rd family PDF, including the top-quark, may play an important role at 100 TeV. This would lead to a process with 2 b-jets in the final state, allowing for a b-tagging on the forward jets, and could  be treated as a different process.}.
The SM parameters used are: $\alpha_{EW}=1/127.9$, $M_Z = 91.18 \, \text{GeV}$, $G_F = 1.16639 \times 10^{-5} \, \text{GeV}$ and $\alpha_S(M_Z) = 0.118$. 
Besides the CH scenario described above, we produced events for the relevant backgrounds: SM EW ZZjj, and the QCD ZZ+jets, merged up to the second jet at LO accuracy through the \MEPS~\cite{Hoeche:2009rj} algorithm as implemented in \Sherpa. 

We would like to stress out the importance of gauge invariance in this study. The cancellations are so delicate that even fixed width effects can produce a large \emph{fake} enhancement at high energies. One way out is to use the complex mass scheme to restore gauge invariance. Our approach is instead to set all the widths of the gauge bosons to zero, since we do not have, due to the implemented generation cuts, kinematic regions where the internal boson propagators go on-shell. Z-bosons are decayed a posteriori with the \Sherpa decay handler.

An analysis routine has been implemented in the \Rivet framework \cite{Buckley:2010ar}. 
Final state particles are identified within $|\eta| < 6$.
One pair of isolated opposite charged muons and one of electrons with $p_{T,\text{min}} = 30 \, \text{GeV}$ and $|\eta_\ell| < 4 $ are identified to reconstruct the $Z$ bosons. If more than one lepton of the same type is present we take the one with highest $p_T$. The reconstructed $Z$ mass is required to be in the window  $65\GeV < m(Z) < 115\GeV$, in order to suppress the non-ZZ backgrounds. 
Jets are reconstructed with the anti-$k_T$ clustering algorithm, with $R=0.4$ and $p_{T,\text{min}} = 30$ GeV. 
Moreover, typical kinematic selection cuts to enhance VBS topology have been implemented for LHC (FCC): 
the two jets are back-to-back in the forward-backward region of the detector forming a system with large invariant mass, while the $Z$-bosons are central and highly energetic.
These cuts are summarized in \tab{tab:cuts}.

\begin{table*}[ht]
\begin{tabular}{|c|c|c|}
\hline
cut & 100 TeV & 14 TeV \\ 
\hline
2 jets & $p_T > 30$ GeV , $|\eta| > 3.5$ , $\eta_1 \cdot \eta_2 < 0$ &  $p_{T,j} > 30$ GeV , $|\eta_j| > 3.$ , $\eta_{j_1} \cdot \eta_{j_2} < 0$\\
ZZ invariant mass & $m_{ZZ} > 3$TeV & $m_{ZZ} > 3$TeV  \\
di-jet invariant mass & $m_{jj} > 1$ TeV & $m_{jj} > 1$ TeV \\
Zs centrality & $|\eta_{Z_i}| < 2.$ & $|\eta_{Z_i}| < 2.$ \\
Zs momentum & $p_{T,Z_i} > 1$ TeV & $p_{T,Z_i} > 0.5$ TeV\\
\hline
\end{tabular}
\caption{Selection cuts implemented in the analyses at the \FCC and \LHC.}
\label{tab:cuts}
\end{table*}

For the statistical assessment we performed a simple counting experiment analysis. We define $S=\sigma_SL$ and $B=\sigma_BL$, where $L$ is the considered integrated luminosity and $\sigma_{S,B}$ are the effective cross sections after the application of all selection cuts for the CH scenario (S) and for the SM prediction (B), both comprizing QCD ZZ+jets. 
We have multiplied the final cross section by a factor 2 assuming the decay channels with 2 pairs of identical leptons can be reconstructed with similar efficiency to the channel $2e2\mu$.
We model the probability to observe a number of events $k$ with a smeared Poisson and mean value $\lambda$, given by either $S$ or $B$,
\begin{equation}
\label{eq:pdfs}
\mathcal{P}(k;\lambda,\epsilon)=\frac{1}{2 \epsilon}
   \int^{1+\epsilon}_{1-\epsilon}\text{d}x\,
     e^{-x \lambda}\frac{(x \lambda)^k}{k!}
\end{equation}
where $\epsilon$ models a flat systematic and theoretical uncertainty, related to scale dependence and experimental systematic error.  

QCD corrections to boson--boson production via vector boson fusion
\cite{Jager:2006zc,Jager:2006cp,Bozzi:2007ur,Jager:2009xx}
at the LHC turn out to be below 10\%. At the FCC this is expected to be even lower. 
EW corrections, on the other hand, are known to increase with energy and can be very large and negative for VBS~\cite{Biedermann:2016yds}\footnote{In $W^\pm W^\pm$ channel at the LHC the EW correction is $k\sim -25\%$ for $M(\ell^\pm\ell^\pm)\gtrsim 500\GeV$ for LHC energies.}.
To partially account for such large corrections we consider a flat error up to $\epsilon=40\%$.

A good estimator of the discriminatory power of the analysis is given by the probability to exclude the SM assuming one of the CH scenarios describes Nature. This probability is given by
\begin{equation}
1-\beta = \sum_{k=m}^\infty \, \mathcal{P}(k;S) 
\label{eq:pbsm}
\end{equation}
where $m$ is defined by
\begin{equation}
\label{eq:ldef}
\sum_{k=0}^m \, \mathcal{P}(k;B) = 95\%\footnote{To ensure exact 95\%  in the formula above we use fractional values in the sum.}.
\end{equation}

\subsubsection{Non-resonant excess at 100 TeV}

We will see below that the non-resonant enhancement cannot be observed at the LHC, we therefore study this scenario at a 100 TeV collider.

\begin{figure*}
\begin{tabular}{c c}
\includegraphics[width=0.5\textwidth,height=0.4\textwidth]{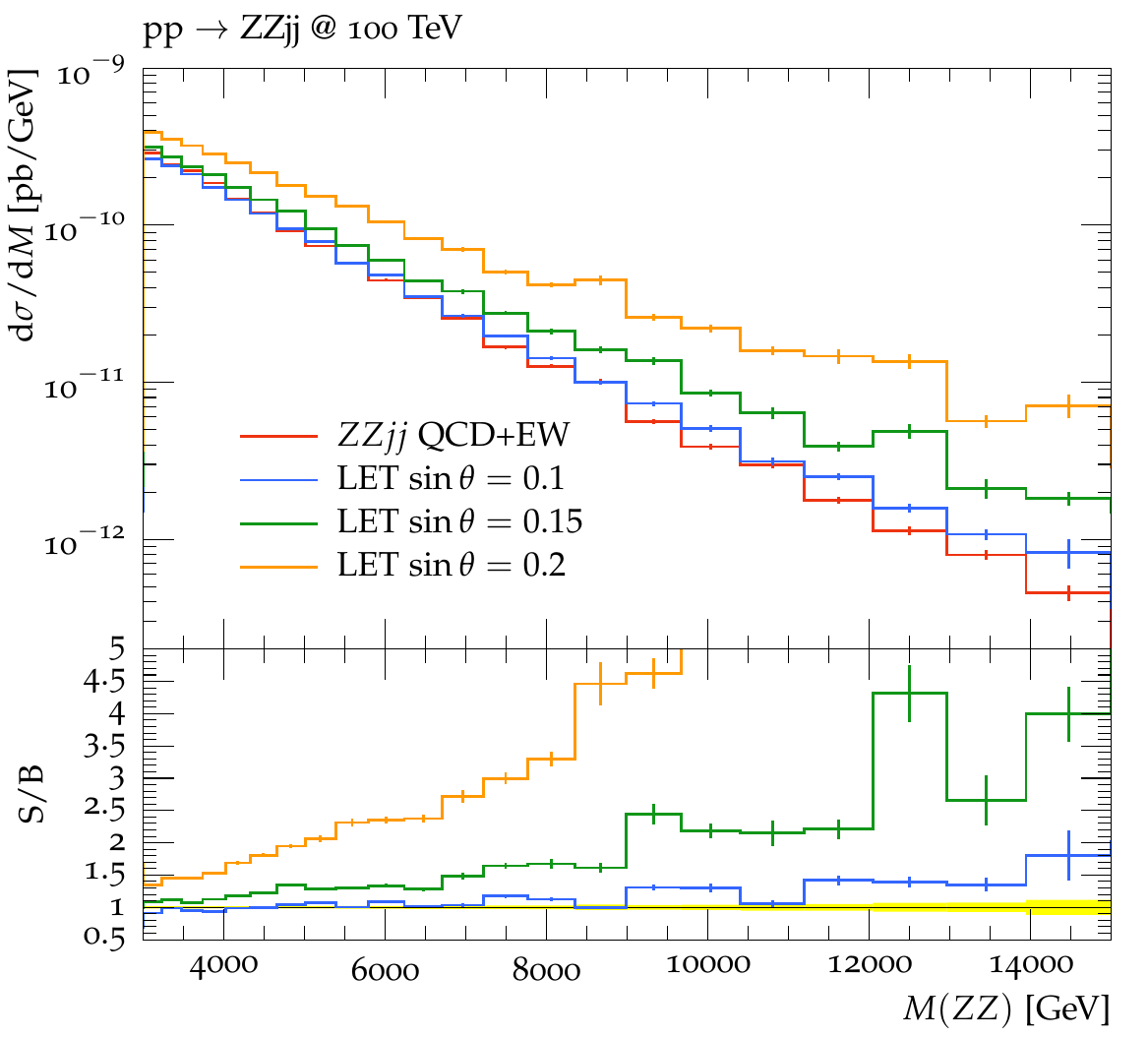} &
\includegraphics[width=0.5\textwidth,height=0.4\textwidth]{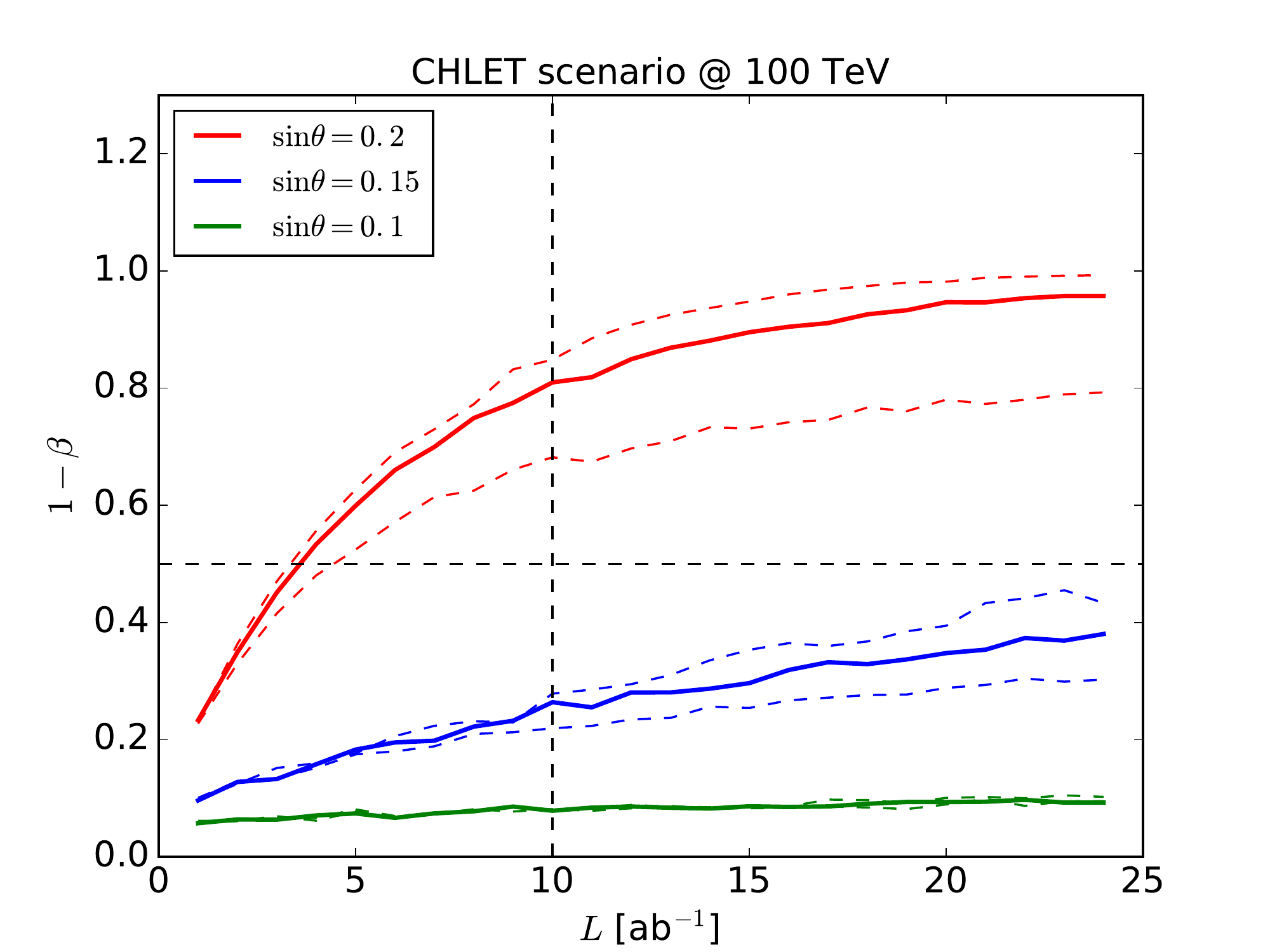} \\
(a) & (b) 
\end{tabular}
\caption{In (a) the ZZ system reconstructed invariant mass distribution for  $\sin\theta=0.1,\,0.15,\,0.2$ in the non-resonant excess scenario and the SM backgrounds (EW ZZjj and QCD ZZ+jets). In  (b) the corresponding $1-\beta$.}
\label{fig:zzmchlet}
\end{figure*}

In \fig{fig:zzmchlet} (\emph{left} panel) we show the distributions of the reconstructed ZZ system invariant mass, for the scenario with $\sin\theta = 0.2 \, , 0.15 \, ,0.1$.
The corresponding $1-\beta$  is shown in the right panel as a function of luminosity, $L$. The central solid line assumes a systematic error $\epsilon=20\%$. The upper and lower dashed lines refer to no-systematic and  $\epsilon=40\%$ respectively. 
The vertical dashed line highlights the benchmark value of luminosity used in the limits set on the vectorial resonances, $L = 10 \, \text{ab}^{-1}$. The line $1-\beta = 0.5$ indicates the exclusion assuming the \emph{mode} of the distribution is observed.

We can see that for the case $\sin\theta=0.2$ we have a good probability ($1-\beta \gtrsim 50\%$) of excluding the SM already around $L\sim 3\iab$. 
For $\sin\theta=0.15$ we need more statistics, with $L\gtrsim 25\iab$ we can reach a good probability to exclude the SM. For $\sin\theta=0.1$ the situation is more complicated and considering the other VBS channels is unavoidable.

\subsubsection{Heavy scalar at 100 TeV}

\begin{figure*}
\begin{tabular}{c c}
\includegraphics[width=0.5\textwidth,height=0.4\textwidth]{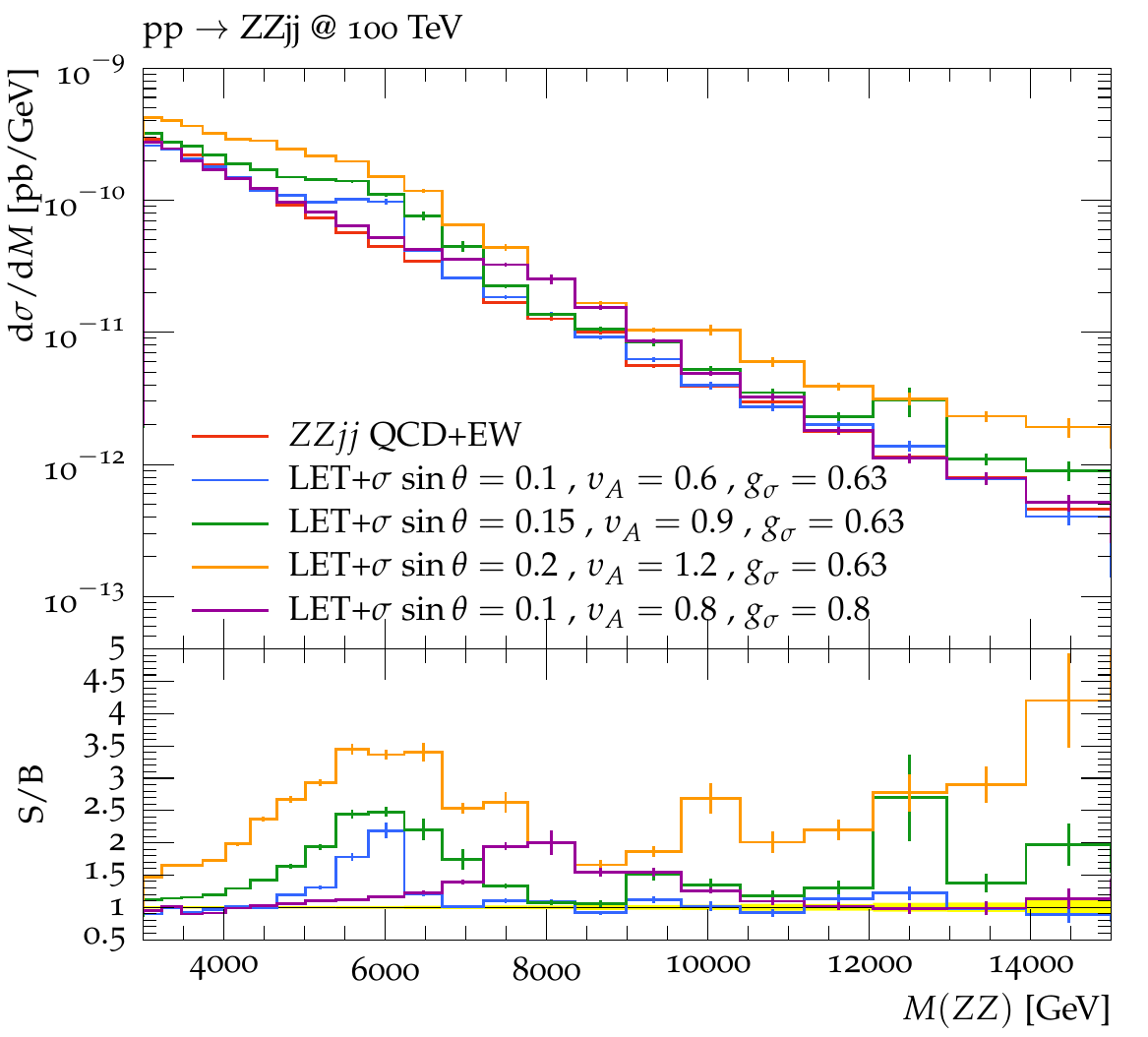} &
\includegraphics[width=0.5\textwidth,height=0.4\textwidth]{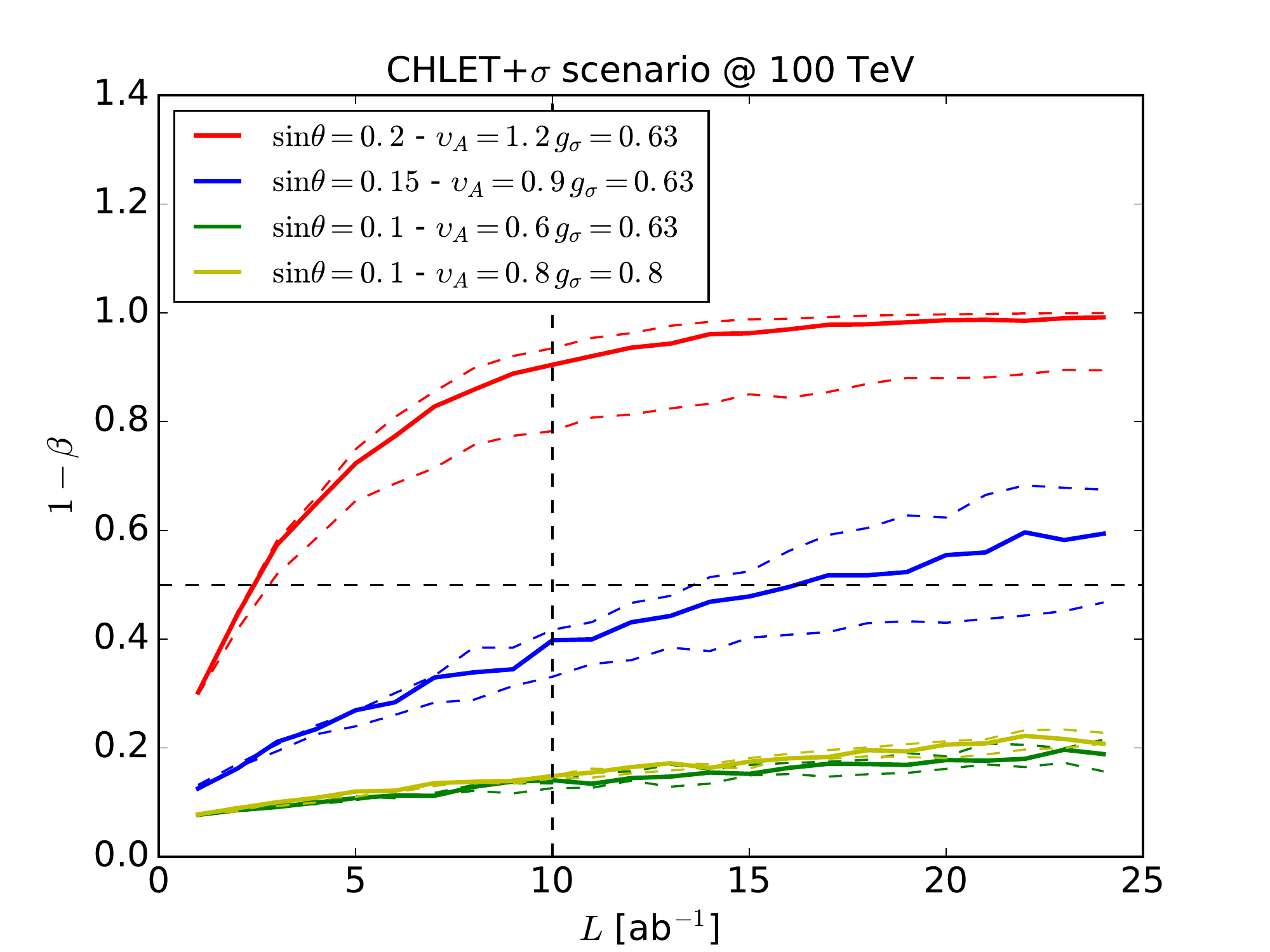} \\
(a) & (b) 
\end{tabular}
\caption{In (a) the ZZ system reconstructed invariant mass distribution in the $\sigma$ resonant excess scenarios (\tab{tab:sigma_benchmark}) and in  (b) the corresponding $1-\beta$.}
\label{fig:zzmsigma}
\end{figure*}

The $\sigma$ resonance has a more pronounced excess at lower energies and a better probability to be observed. 
In \fig{fig:zzmsigma} (a) we present the invariant mass of the reconstructed ZZ system for the resonant scenarios listed in \tab{tab:sigma_benchmark}. 
We note that the $\sigma$ resonance postpones the unitarity violation with respect to the plain LET scenario, and the high energy behavior beyond the resonance peak approaches the SM prediction for a large energy range. For this reason we add a selection $M(ZZ)<10\TeV$ to avoid contamination from non-resonant areas. 

In \fig{fig:zzmsigma} (b) the corresponding $1-\beta$ are shown. 
We note a good probability  $1-\beta>$ 50\% even for $\sin\theta=0.15$, which could be in particular stronger than vector resonance searches. 

\subsubsection{Heavy scalar at the \LHC}

The LHC is not the most indicated machine to observe signal of strong VBS in CH models due to the high intrinsic compositeness scales. However, nothing prevents that some dynamical mechanism produces a lighter state. 

BSM searches through VBS have been analysed by ATLAS and CMS collaborations~\cite{ATL-PHYS-PUB-2013-006,ATLAS-CONF-2016-112,Khachatryan:2014sta}. 
In \cite{ATLAS-Collaboration:1496527} in particular the production of scalar resonances in VBS in the $ZZ\to 4\ell$ channel at $\sqrt{s} = 14 \TeV$, for $L = 300 - 3000\ifb$ has been considered. For a resonance of mass $M_{\sigma} = 1\TeV$ with $g_{\sigma} = 2.5$ they predict 
a sensitivity of $9.4$ standard deviations at $3 \, \text{ab}^{-1}$.
Unfortunately, our motivated scenarios have masses larger and smaller couplings.
We consider here  $g_\sigma=0.8$, $M_\sigma=4\TeV$ as an optimistic case.

In \fig{fig:lhc_masses} we show the invariant mass of the reconstructed $ZZ$ system  at  $\sqrt{s} =14 \TeV$. 
The effective cross section found is only $\sigma= 2.9 \times 10^{-4} \, \text{ab}$.
As already noted, the $ZZ$ channel has the smallest cross-section amongst the VBS channels and including all the other channels is imperative for this search. 
Another source of improvement could come from the mixing of $\sigma$ with the Higgs, which at this mass could give some small gluon fusion contribution. 
Further more detailed study is required. 

\begin{figure}
\includegraphics[width=0.5\textwidth]{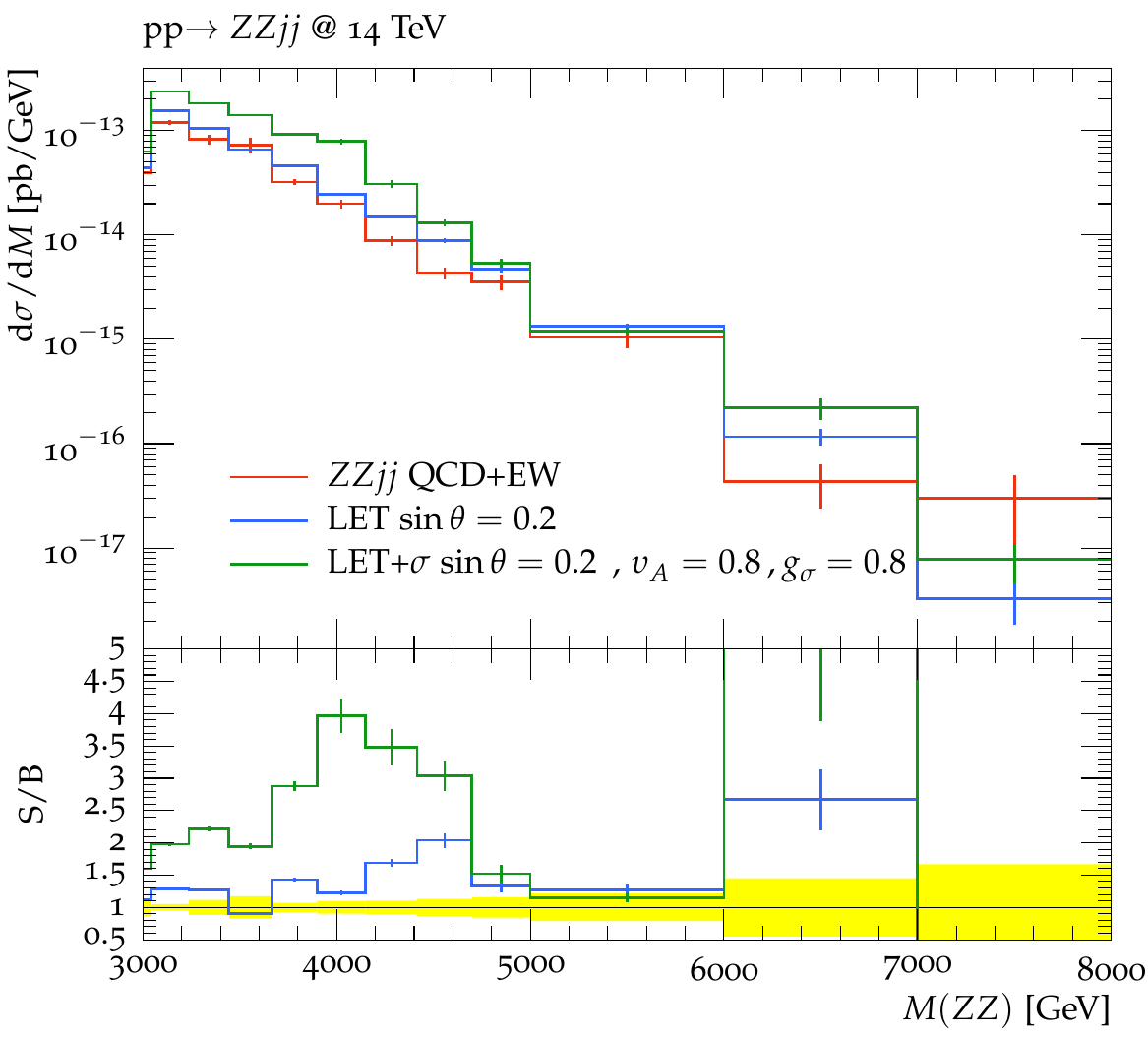}
\caption{ZZ invariant mass at the \LHC for the composite scenario devised for LHC ($g_\sigma=0.8$, $M_\sigma=4\TeV$) and the SM backgrounds (EW ZZjj and QCD ZZ+jets). }
\label{fig:lhc_masses}
\end{figure}

\section{Conclusions}
\label{sec:conclusion}

In this work we have shown the implications of GBS unitarity in the spectra of CH scenarios, in particular the FMCHM. 

We have especially made definite predictions for the possible range of the mass of an eventual $\sigma$-like composite scalar resonance, which can be described as a Breit-Wigner peak only if $M_\sigma\lesssim 1.2\TeV/\sin\theta$. Heavier than that, the LET growing behavior overcomes and dilutes any possible peak, making the strong VBS signal more like a \emph{continuum}.

Inspired by models of unitarization, which saturate unitarity and provide good description of pion-pion and pion-kaon scattering data, we estimate the parameters of the  CH effective description.
The vector and $\sigma$ couplings to NGBs are found to be $a_V\sim 1$ and  $g_\sigma\lesssim 0.63$. 

The predictions from the analysis of GBS amplitudes lead to specific signatures at experimental set-ups in colliders.
Limits on the production cross section of heavy composite vector  in the FMCHM   and a first assess of strong VBS in CH in the simplest channel $pp\to jjZZ\to jj4\ell$ have been provided. The results at the 100 TeV collider are promising. 
For $\sin\theta=0.2$ the non-resonant excess could be observed with few inverse attobarns
and an indirect limit on $hVV$ coupling set.
The scenario $\sin\theta=0.15$ will also be observed with modest luminosities $L\sim 30\iab$. 
The motivated resonant scenario with $\sin\theta=0.15$ and $g_\sigma=0.63$ would probably be detected with 
$L\sim 15\iab$. 
For lower values of $\sin\theta$ other VBS channels  must be considered to enhance the observability potential.  
At the LHC the situation is more complicated even for very optimistic scenarios, and a more detailed study including other VBS channels and gluon fusion production must be considered.


\section*{Acknowledgements}

We thank Steffen Schumann for careful reading of the manuscript and useful discussions. DBF thank Maria J. Herrero for discussions about unitarization models.

\appendix


\section{FMCHM in the CCWZ construction }
\label{app:ccwz}

We follow Refs.~\cite{Galloway:2010bp,Cacciapaglia:2014uja} for the symmetry breaking structure of the theory. 
The unbroken $V^a$ ($a=1\cdots 10$) and spontaneously broken $Y^a$ ($a=1 \cdots 5$) generators in SU(4)/Sp(4) are defined by
\begin{eqnarray}
V^a \cdot \Sigma_0 + \Sigma_0 \cdot {V^a}^T &=& 0\,,\\ 
Y^a \cdot \Sigma_0 - \Sigma_0 \cdot {Y^a}^T &=& 0\,.
\label{eq:VY}
\end{eqnarray}
We can similarly define the unbroken and broken generators in the $\Sigma_B$ vacuum,
\begin{eqnarray}
S^a \cdot \Sigma_B + \Sigma_B \cdot {S^a}^T &=& 0\,,\\ 
X^a \cdot \Sigma_B - \Sigma_B \cdot {X^a}^T &=& 0\,,
\end{eqnarray}
and identify $S^i$ (i=1,2,3) and $S^i$ (i=4,5,6) as the  SU(2)$_L$ and SU(2)$_R$ subgroups generators.
SU(2)$_L$ and $Y\equiv S_6$ are then gauged and identified as the EW group, weak and hypercharge respectively. 
When $\theta$ is non-zero, the unbroken generators $V^a$ are not fully aligned with the electroweak generators and EW symmetry is spontaneously broken.

The (pseudo-)NGB field is parametrized by the exponential map
\beq
U = \exp \left[ \frac{i\sqrt{2}}{f} \sum_{a=1}^5 \pi^{a} Y^a \right]   \,,
\eeq
with $\pi_{a}$ the NGBs. $\pi^4\equiv h$ is identified with the Higgs boson and $\pi^5\equiv \eta$ a electroweak singlet, and they are pseudo-NGBs. The other 3 are the exact NGBs absorbed by $W$ and $Z$ bosons. 

We define the gauge Maurer-Cartan one-form $\omega_{\mu}$ and its projections,
\beq
& \omega_{\mu} &= U^{\dagger} D_{\mu} U \text{,} \\
& D_{\mu} &= \partial_{\mu} -i g W^i_{\mu} S^i - i g' B_{\mu} S^6 , \\ 
& x_{\mu} &= 2 \tr \left[ Y_a \omega_{\mu} \right] Y^a \ ,\\
& s_{\mu} &= 2 \tr \left[ V_a \omega_{\mu} \right] V^a \ .
\label{eq:proj}
\eeq
$v_{\mu}$ transforms inhomogeneously under $SU(4)$
\begin{equation}
v_{\mu}\to v_{\mu}^\prime=h(g,\pi)\, (v_{\mu}+i\partial_\mu)\, h^\dagger(g,\pi)\,,
\end{equation}
while $x_{\mu}$ transforms homogeneously
\begin{equation}
x_{\mu}\to x_{\mu}^\prime=h(g,\pi)\, x_{\mu}\, h^\dagger(g,\pi)\,.
\end{equation}

\subsection{Vector resonances in the HLS}

In the HLS method, we enhance the symmetry group 
${\mathrm{SU(4)}}$ to ${\mathrm{SU(4)}}_0\times {\mathrm{SU(4)}}_1$, and embed the SM gauge bosons in SU(4)$_0$ and the heavy resonances in SU(4)$_1$.
The low energy Lagrangian is then characterised in terms of the breaking of the extended symmetry down to a single Sp(4):
the ${\mathrm{SU(4)}}_i$ are spontaneously broken to 
${\mathrm{Sp(4)}}_i$ via the introduction of 2 matrices $U_i$ containing 5 NGBs each. The remaining ${\mathrm{Sp(4)}}_0\times {\mathrm{Sp(4)}}_1$ is then spontaneously broken to ${\mathrm{Sp(4)}}$ 
by a sigma field $K$, containing 10 NGBs corresponding to the generators of Sp(4). 
The two replicas of the NGB exponential map are given by
\begin{eqnarray}
	U_0 &=& \exp\left[ \frac{i \sqrt 2}{f_0} \sum_{a=1}^5 ( \pi_0^a  Y^a )  \right],   \nonumber \\
    U_1 &=& \exp\left[ \frac{i \sqrt 2}{f_1} \sum_{a=1}^5 ( \pi_1^a  Y^a )  \right]\,.
\end{eqnarray}
The Maurer-Cartan one form and its projections to the broken generators are defined for each copy in the same way as in in \eq{eq:proj}, with the weak bosons and the heavy composite vectors introduced in the different copies of $U_i$, as
\begin{eqnarray}
D_\mu U_0 &=& ( \partial_{\mu} -i g W^i_{\mu} S^i - i g' B_{\mu} S^6 ) U_0\,,  \nonumber\\
D_\mu U_1 &=& (\partial_{\mu} -i \widetilde{g} {\cal V}_\mu^a V^a - i \widetilde{g} {\cal A}_\mu^b Y^b ) U_1 \,.
\end{eqnarray}

The $K$ field is introduced to break the two remaining copies of $Sp(4)$, $Sp(4)_0\times Sp(4)_1$ to the diagonal final $Sp(4)$:
\begin{equation}
	K = \exp\left[ i k^a V^a / f_K \right]\,,
\end{equation}
and it transforms like
\begin{equation}
K\to K^\prime=h(g_0,\pi_0)\, K\, h^\dagger(g_1,\pi_1)\,,
\end{equation}  
thus its covariant derivative takes the form
\begin{equation}
D_\mu K = \partial_\mu K -i v_{0\mu} K + i K v_{1\mu} \,.
\end{equation}  
The 10 pions contained in $K$ are needed to provide the longitudinal degrees of freedom for the 10 vectors $\mathcal{V}^a_\mu$, while a combination of the other pions $\pi^i$ acts as the longitudinal degrees of freedom for the $\mathcal{A}^a_\mu$. It should be reminded that out of the 5 remaining scalars, 3 are exact NGBs \emph{eaten} by the massive $W$ and $Z$ bosons, while 2 remain as physical scalars in the spectrum: one Higgs-like state plus a singlet $\eta$.

In charge eigenstate, the trilinear couplings (\eq{eq:vtril})  are given by:
\begin{eqnarray}
\pi^+\pi^-v^0 &:&  i g_V (p_- - p_+)\,, \label{eq:vgcoup1}\\ 
\pi^\pm\pi^0 v^\mp&:&   i g_V (p_\pm - p_0)\,,  \\
\pi^\pm h s^\mp &:& g_V (p_\pm - p_h) \,, \\ 
\pi^0 h s^0&:&  g_V (p_0 - p_h)  \,,\\
\pi^\pm \eta \tilde{s}^\mp &:&    \mp i g_V (p_\pm - p_\eta)  \,, \\ 
 \pi^0 \eta \tilde{s}^0&:&   i g_V (p_0 - p_\eta) \,,\\
h \eta \tilde{v}^0 &:& - g_V (p_h - p_\eta) \label{eq:vgcoup2}\,.
\end{eqnarray}
We have used a redefinition $\tilde{s}^0\to -i\tilde{s}^0$ w.r.t. Ref.\cite{Franzosi:2016aoo}.



\section{Partial waves and inelastic cross section}
\label{app:pw}

One can expand the $\pi\pi\to\pi\pi$ scattering amplitudes in partial waves according to \eq{eq:pw}.
The elastic differential cross section is given by (neglecting external masses and for identical particles)
\begin{equation}
d\sigma_{el} = \frac{|A(s,t)|^2}{64\pi\,s}d\cos\theta\,.
\end{equation}
By integrating in $\cos\theta$ and using the completeness relation of Legendre polynomials we get the contribution from the elastic channel to the cross section,
\begin{equation}
\sigma_{el} = \frac{32\pi}{s}\sum_J (2J+1)|a_J(s)|^2\,.
\end{equation}
Now we can get the total cross section by the use of the optical theorem
\begin{equation}
\sigma = \frac{1}{s}\Im A(s,0) = \frac{32\pi}{s}\sum_{J=0}^\infty\imag a_J(s)(2J+1)P_J(1)\,,
\end{equation}
and derive an expression for the total cross section of inelastic channels $\pi\pi\to X$, which must account for the rest of the total cross section. Hence,
\begin{equation}
\sigma_{inel}=\sigma-\sigma_{el}= \frac{32\pi}{s}\sum_{J=0}^\infty (2J+1) \left(\imag a_J(s) - |a_J(s)|^2 \right)\,.
\end{equation}
We can now define
\begin{equation}
\sum_X|a_J^{X}(s)|^2\sqrt{1-M_X^2/s}\equiv \imag a_J(s) - |a_J(s)|^2\,,
\end{equation}
where $X$ is one inelastic channel and $M_X^2$ its total squared mass.
This function has a maximum $\sum_X|a_J^{X}(s)|^2\sqrt{1-M_X^2/s}<1/4$, which is related to the Froissart bound for inelastic channel~\cite{Martin:2015kaa}.


\bibliographystyle{apsrev4-1}

\bibliography{FCHVBS}

\end{document}